%% file: document.tex
\documentclass[twocolumn]{aastex63}

\usepackage{rotating}
\usepackage{amssymb}
\usepackage{amsfonts}
\usepackage{amsmath}
\usepackage[varg]{txfonts}
\usepackage{natbib}
\usepackage{color}
\usepackage{comment}
\usepackage{graphicx,rotating}

\usepackage{longtable}
\usepackage{threeparttablex}
\usepackage{lineno}

\usepackage{url}
\usepackage{float}
\usepackage{hyperref}
\usepackage{multirow}

\usepackage{graphicx}
\usepackage{subfigure} 
\usepackage{xtab}
\usepackage{amsmath} 
\usepackage{amssymb} 

\usepackage{graphicx}
\usepackage{array} 
\usepackage{multirow}
\usepackage{threeparttable}
\usepackage{tabularx}

\graphicspath{/}
\hypersetup{colorlinks, linkcolor = {blue} }

\received{\today}
\revised{---}
\accepted{---}
\submitjournal{--- Journal}

\shorttitle{X.Y Ma (2021) A new pulsating sdB}

\shortauthors{Ma et al.(2021)}

\begin{document}
	
\title{K2 photometry on oscillation mode variability: the new pulsating hot B subdwarf star EPIC 220422705}



\author{Xiao-Yu~Ma}
\affiliation{Department of Astronomy, Beijing Normal University, Beijing 100875, People’s Republic of China}

\author{Weikai~Zong}
\affiliation{Department of Astronomy, Beijing Normal University, Beijing 100875, People’s Republic of China}
\email{weikai.zong@bnu.edu.cn; jnfu@bnu.edu.cn}

\author{Jian-Ning~Fu}
\affiliation{Department of Astronomy, Beijing Normal University, Beijing 100875, People’s Republic of China}

\author{M. D. Reed}
\affiliation{Department of Physics, Astronomy and Materials Science, Missouri State University, 901 S. National, 65897 Springfield, MO,USA}

\author{Jiaxin Wang}
\affiliation{Department of Astronomy, Beijing Normal University, Beijing 100875, People’s Republic of China}
\affiliation{College of Science, Chongqing University of Posts and Telecommunications, Chongqing 400065, People's Republic of China}

\author{St\'ephane Charpinet}
\affiliation{Institut de Recherche en Astrophysique et Plan\'{e}tologie,~CNRS,~Universit\'{e} de Toulouse, CNES,~14 Avenue Edouard Belin,~31400 Toulouse,~France}

\author{Jie Su}
\affiliation{Yunnan Observatories, Chinese Academy of Sciences, Kunming 650216, People’s Republic of China}
\affiliation{Key Laboratory for the Structure and Evolution of Celestial Objects, Chinese Academy of Sciences, Kunming 650216, People’s Republic of China}

\begin{abstract}
We present analysis of oscillation mode variability in the hot B subdwarf star EPIC~220422705, a new pulsator discovered from $\sim78$~days of {\em K}2 photometry. The high-quality light curves provide a detection of 66 significant independent frequencies, from which we identified 9 incomplete potential triplets and 3 quintuplets. Those {\sl g-} and {\sl p-}multiplets give rotation periods of $\sim$ 36 and 29 days in the core and at the surface, respectively, potentially suggesting a slightly differential rotation. We derived a period spacing of 268.5\,s and 159.4\,s for the sequence of dipole and quadruple modes, respectively. We characterized the precise patterns of amplitude and frequency modulations (AM and FM) of 22 frequencies with high enough amplitude for our science. Many of them exhibit intrinsic and periodic patterns of AM and FM, with periods on a timescale of months as derived by the best fitting and \texttt{MCMC} test. The nonlinear resonant mode interactions could be a natural interpretation for such AMs and FMs after other mechanisms are ruled out. Our results are the first step to build a bridge between mode variability from {\em K}2 photometry and nonlinear perturbation theory of stellar oscillation.
\end{abstract}


\section{Introduction} \label{sec:intro}
Hot B subdwarf (sdB) stars are burning helium in the core and typically wrapped in a thin hydrogen envelop at the surface. Their compact ($\log g = 5.2 \sim 6.2$~dex) and hot ($T_{\rm{eff}} = 20000 \sim 40000$\,K) properties place them to the extreme horizontal branch (EHB) in the Hertzsprung–Russell diagram \citep[see][for a review]{2009ARA&A..47..211H,2016PASP..128h2001H}. A fraction of those blue faint objects have luminosity variations which can be attributed to oscillations of gravity ($g$-) or pressure ($p$-) modes or both \citep{2003ApJ...583L..31G,1997MNRAS.285..640K,2006A&A...445L..31S}. Those modes are driven by the classical $\kappa$-mechanism due to an opacity bump produced by ionization of iron group elements \citep{1996ApJ...471L.103C,1997ApJ...483L.123C,2003ApJ...597..518F}. 
Due to their rich oscillations, sdB variables (sdBV) are good candidates to probe their interior via the tool of asteroseismology \citep{2005A&A...443..251C}.

As advanced by observations from space, for instance, {\sl Kepler/K}2 and TESS \citep{2010Sci...327..977B,2014PASP..126..398H,2015JATIS...1a4003R}, oscillation frequencies in sdBV stars can be sharply resolved to unprecedented high precision, which leads to fruitful achievements for probing the interior of sdB stars \citep[see, e.g.,][]{2019A&A...632A..90C,2014MNRAS.440.3809R,2010ApJ...718L..97V}. There are 18 sdBV stars discovered in the original {\em Kepler} field \citep{2010MNRAS.409.1470O,2011MNRAS.414.2860O,2011ApJ...740L..47P,2012MNRAS.427.1245R}, among which most stars had been continuously observed after they were discovered to pulsate. In  contrast to ground-based photometry, several sdBV stars are found with more than 100 frequencies such as  KIC\,03527751 \citep{2015ApJ...805...94F,2018ApJ...853...98Z}. A preliminary mass survey on sdBV stars established that they are distributed around the canonical value $\sim0.47 M_{\odot}$ in a narrow region \citep{
2012A&A...539A..12F} with rotational periods distributed from a few days up to even hundreds of days \citep{2018OAst...27..112C,2021MNRAS.500.2461S,2021MNRAS.507.4178R}. In individual analyses, many sdBVs show clear variations in amplitude with a timescale much longer than their oscillation periods \citep{2014MNRAS.440.3809R,2016A&A...594A..46Z,2017MNRAS.465.1057K}.
Focusing on amplitude modulations, \citet{2016A&A...594A..46Z} found that frequencies are not stable for many rotational components in KIC\,10139564. They concluded that the amplitude and frequency modulations (AM/FM) can be attributed to nonlinear interactions of resonant mode coupling \citep{1994A&A...291..481G,1995A&A...296..405B,1997A&A...321..159B}, a mechanism of intense focus in other pulsators, for instance pulsating white dwarfs \citep{2016A&A...585A..22Z} and slowly pulsating B stars \citep{2021A&A...655A..59V}. Observational AM/FM variations provide strong constraints for the development of nonlinear stellar oscillation theory.

However, the {\em Kepler} space telescope had to begin the reborn mission with pointing using only two reaction wheels. This so-called {\em K}2 phase provided nearly-uninterrupted photometry for almost three months but could observe a larger spatial coverage than the original {\em Kepler} mission. Therefore, {\em K}2 offers a higher chance to finding more sdBV stars. In the 20 campaigns of {\em K}2,  nearly 200 sdBV candidates were observed to search for pulsations or transits, leading to 10 sdBV stars already published \citep[see, e.g., ][]{2019MNRAS.483.2282R,2019MNRAS.489.1556B,2019MNRAS.489.4791S}. These $\sim 80$ d {\em K}2 observations could also be helpful to characterize the amplitude modulations of pulsation modes in sdBVs \citep[see, e.g.,][]{2019MNRAS.489.4791S}. Similar to {\em Kepler} results, {\em K}2 photometry will shed new light on AM/FM oscillations in sdBV stars on shorter-term timescales.  

As demonstrated by a series of works from  \citet{2016A&A...594A..46Z,2016A&A...585A..22Z,2018ApJ...853...98Z}, evolved compact pulsators, including pulsating white dwarfs and sdBVs, could be excellent candidates to provide observational constraints to develop the nonlinear amplitude equations which describe how amplitudes and frequencies modulate. Gained from those experiences, we initiated a new survey of AM/FM in sdBV stars from {\em K}2, on relatively shorter modulation timescales appropriate for {\em K}2. In this paper, we concentrate on the bright sdB star, EPIC~220422705, or PG~0039+049, which has $K_p$ = 12.875 and is located at $ \rm \alpha = 00{^h}42{^m}06{^s}.124$  and $\rm \delta = +05{^d}09{^m}{23{^s}.376}$. This star was originally identified as a faint blue star by \cite{1980A&AS...39...39B} and then was classified as an sdB star with spectra \citep{1988SAAOC..12....1K}.  \citet{1990A&A...239..265M} derived atmospheric parameters of $T_\mathrm{eff} = 26700$\,K and $\log g = 4.7$~dex for EPIC~220422705,  with a distance of $d = 1050\pm400$~pc and refined by GAIA EDR3 to $d = 916.6^{+69.2}_{-81.5}$\,pc  \citep{2021A&A...649A...1G}. It is a binary system containing a cool companion as disclosed by \citet{2011MNRAS.415.1381C} and further confirmed as a G2V dwarf star with a preliminary period of $150–300~{(\pm220)}$ days \citep{2012ApJ...758...58B}. 
The structure of the paper is organised as follows: we analyze the photometric data from {\em K}2 and analyze the asteroseismic properties in Sect.~\ref{sec-data}. We then characterized the amplitude and frequency modulations of 22 frequencies in Sect.~\ref{sec:AF}, followed by a discussion of those modulation details in Sect.~\ref{discuss}. Finally, we summarize our findings in Sect.~\ref{sec:Dis-Conc}.

\section{Frequency content}\label{sec-data}
\subsection{Photometry and frequency extraction}

\startlongtable
\begin{deluxetable*}{cccccccccccc}
    \label{Tab:freq}
	\input{Fre_n_l}
\end{deluxetable*}

\begin{figure}
	\centering 
	\includegraphics[width=8.4cm]{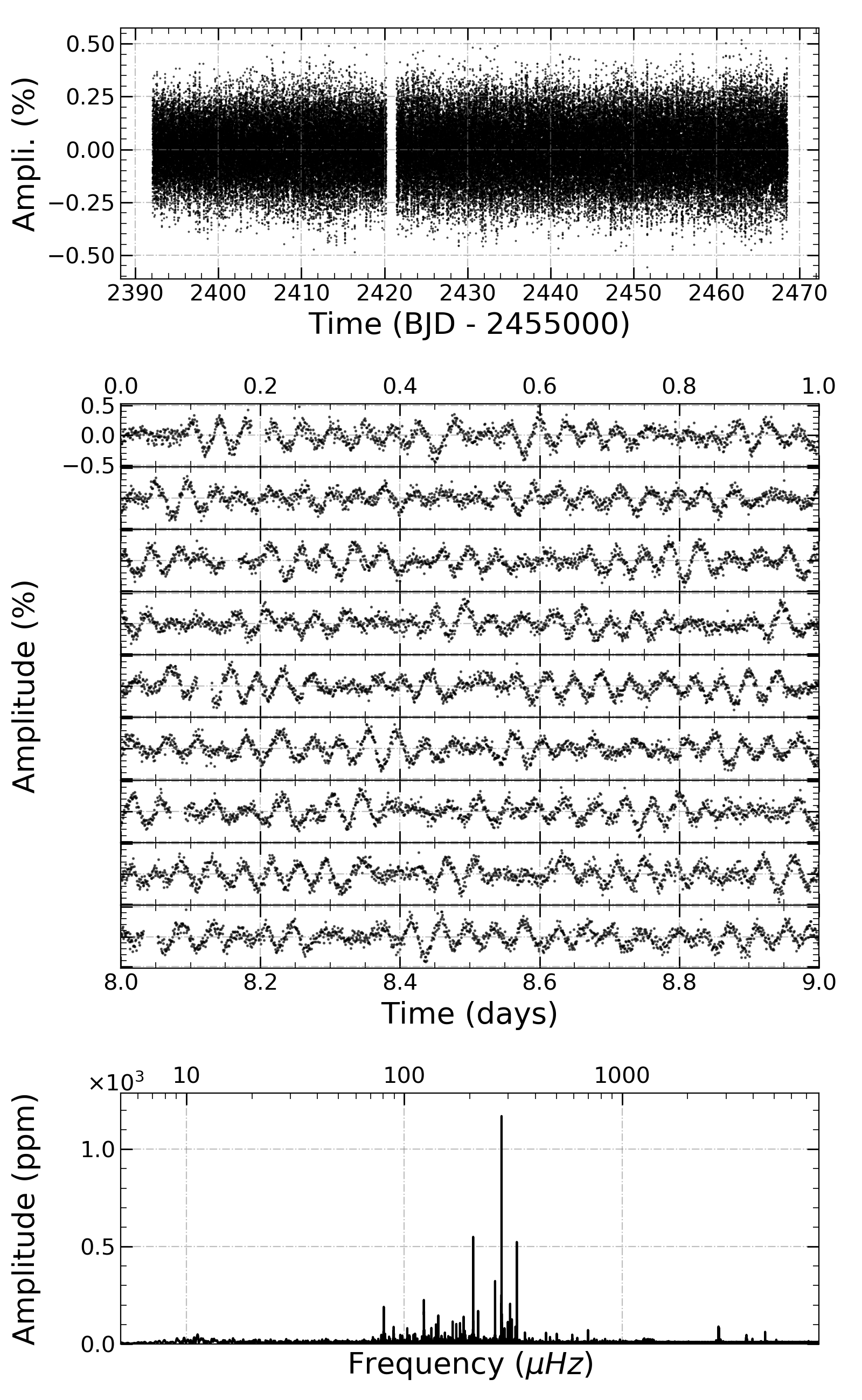}
	\caption{ {\em K}2 photdedao d ometry and frequency signals obtained for EPIC~220422705. \emph{Top panel}: the complete light curve (amplitude is in  percentage, a.k.a. \%, of the mean brightness) with a data sampling of 58.85~s.
	\emph{Middle panel}: a close-up view of a 9-d light curve (starting at BJD~2457433) with each panel having a one-day slice.
	\emph{Bottom panel}: the Lomb-Scargle periodogram of the assembled light curve (amplitude in ppt vs frequency in $\mu$Hz on a logarithmic scale). 
    }
	\label{light-curve}
\end{figure}

EPIC~220422705 had been observed by {\em K}2 in short-cadence (SC) mode over a period of 78.72~days during Campaign~8. 
Its assembled light curves were downloaded from Mikulski Archive for Space Telescopes\footnote{\url{https://archive.stsci.edu/k2}} (MAST).
These archived data were processed through the EVEREST pipeline\footnote{The EPIC Variability Extraction and Removal for Exoplanet Science Targets as developed by Dr. R.  Luger which is an open-source pipeline for removing $\sim 6.5$~hr instrumental systematics in {\em K}2 light curves. One can see details through the link: \url{https://archive.stsci.edu/hlsp/everest}.}. The photometry corrected by EVEREST has  comparable precision to the original {\em Kepler} mission for targets brighter than $K_p \approx 13$ \citep{2018AJ....156...99L}. EPIC~220422705 is within this brightness range.

The EVEREST flux was firstly shifted to the relative fraction to their mean value. We then used a six-order polynomial fitting to detrend the { whole} light curve due to residual instrumental drifts.
To avoid discontinuities in the light curve across gaps longer than { 0.02} days, we separated the light curve piece-wise where such gaps occurred for our fitting. { This detrending method will flatten the light curves and dismiss signals with period ($\gtrsim 2$~d) in Fourier transform, which will not have impact on the modulating patterns for our prime aim. We note that those signals are not concerned here due to the fact that the EVEREST pipeline may not recover those signals correctly.} Then the light curves were iteratively clipped of a few outliers three times by filtering at 4.5$\sigma$ around the light curve before we produced a Fourier transformation. Figure \ref{light-curve} (top panel) shows the final light curve of EPIC~220422705 which contains 106,444 data points over a duration of 76.43\,days with a 1.2~d gap in the middle.
The amplitude scatter clearly reveals multiperiodic signals of hours in a close-up view (middle panel). The corresponding Lomb-Scargle periodogram \citep[LSP;][]{1976Ap&SS..39..447L,1982ApJ...263..835S} up to the Nyquist frequency is shown in the bottom panel where the {\sl g}-mode frequencies are clearly dominant in a region of [$\sim 100 - 1000$]~$\mu$Hz.

We used the specialized software \texttt{FELIX}\footnote{Frequency Extraction for Lightcurve exploitation, developed by S.~Charpinet, greatly optimizes the algorithm and accelerates the speed of calculation when performing frequency extraction from dedicated consecutive light curves. See details in \citet{2010A&A...516L...6C,2019A&A...632A..90C} and \citet{2016A&A...585A..22Z,2016A&A...594A..46Z}.} to perform frequency extraction from the light curves. The frequencies were prewhitened in order of decreasing amplitude until the value of 5.2 times the local noise level, a value that is the median amplitude in the LSP \citep{2021ApJ...921...37Z}. This detection threshold is adopted as a compromise between 2-yr {\em Kepler} and 27-d TESS photometry \citep{2016A&A...585A..22Z,2019A&A...632A..90C}. The highest peak will be extracted in the case where there are several close frequencies of $< 0.4\,\mu$Hz, i.e., about $3\times \Delta f$ ($\Delta f = 1/T$, and T $\sim76.43$ days). We have detected 66 independent frequencies and 13 linear combination frequenciess, with the highest (1172~ppm) frequency at 279.767\,$\mu$Hz, which are listed in Table~\ref{Tab:freq}.

\subsection{p- and g-modes}

\begin{figure}
	\centering 
    \includegraphics[width=8.4cm]{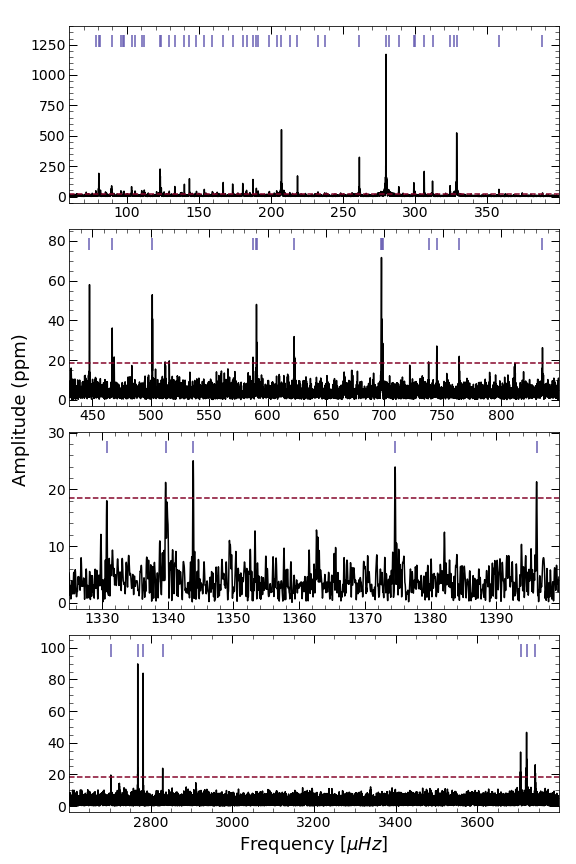}
    
	\caption{Close-up views of the LSP of EPIC~220422705. The entire  periodogram is divided into two different ranges: the low frequency {\sl g}-mode region (two top panels) and the high frequency {\sl p}-mode region (two bottom panels).
    The horizontal dashed line denotes the 5.2$\sigma$ detection threshold and the (blue) vertical segments at the top of each panel indicates locations of extracted frequencies.
	}
	\label{fig:fre_4}
\end{figure}

Pulsating stars with both acoustic {\sl p}- and gravity {\sl g}-mode oscillations are excellent candidates to probe their internal profiles since acoustic and gravity waves propagate in different regions of the stellar structure \citep{ 2010aste.book.....A,2022arXiv220111629K}. From spaceborne photometry, some {\sl g}-mode dominated sdBVs are found with low-amplitude {\sl p}-mode pulsations \citep[see, e.g.,][]{2017A&A...597A..95B,2018ApJ...853...98Z,2020MNRAS.495.2844S}.  A direct and easy way to distinguish the two different types of mode is by their pulsation period.  In general, theoretical sdB star calculations suggest that dipole {\sl p}-modes typically have frequency $>2500$\,$\mu$Hz ( $P<400$\,s), whereas {\sl g}-modes $<1000$\,$\mu$Hz ($P >1000$\,s) \citep{2003ApJ...597..518F,2005A&A...443..251C,2011A&A...530A...3C}.  But p-mode frequencies can decrease below 1700~$\mu$Hz (periods can increase beyond 600~s) as $T_\mathrm{eff}$ and $\log g$ decreases \citep{1999PhDT........26C,2001PASP..113..775C,2002ApJS..139..487C}.

Figure\,\ref{fig:fre_4} shows preliminary classification for the {\sl p}- and {\sl g}-mode regions based only on the period. We detect 53 independent frequencies in the range [$\sim80-1000$]~$\mu$Hz which are clearly {\sl g}-mode ($P>1000$~s) pulsations (two top panels). Another 7 independent frequencies are found in the high frequency {\sl p}-mode ($P<400$~s) region, [$\sim2600-3800$]~$\mu$Hz (bottom panel). There are six independent frequenciesin the region of [$\sim1100-1400$]~$\mu$Hz or  [$\sim715-910$]~s, which might be low-order high-degree ( $\ell>3$) {\sl g}-modes  or mixed modes that need further classification \citep[see, e.g.,][]{2011A&A...530A...3C,2019A&A...632A..90C}. { Those frequencies, hardly directly classified to be {\sl p}- or {\sl g}-mode by merely of their frequency value, can be used to penetrate a much larger portion of stellar interior or to detect the differential rotation in radial or longitude.} However, determining the exact modes requires an exploration of seismic models. We note that a few frequencies were detected in this intermediate region in sdB stars observed with {\em Kepler} photometry, for instance, KIC~3527751 and KIC~10001893 \citep{2015ApJ...805...94F,2017MNRAS.472..700U}. In addition, we have resolved 13 linear combinations with frequencies $<1400~\mu$Hz which could be intrinsic resonant modes \citep{2016A&A...594A..46Z} or non-linear effects from the linear eigenfrequencies \citep{1995ApJS...96..545B}.

\subsection{Rotational multiplets}
\begin{figure*}
    \centering
    \includegraphics[width=\textwidth]{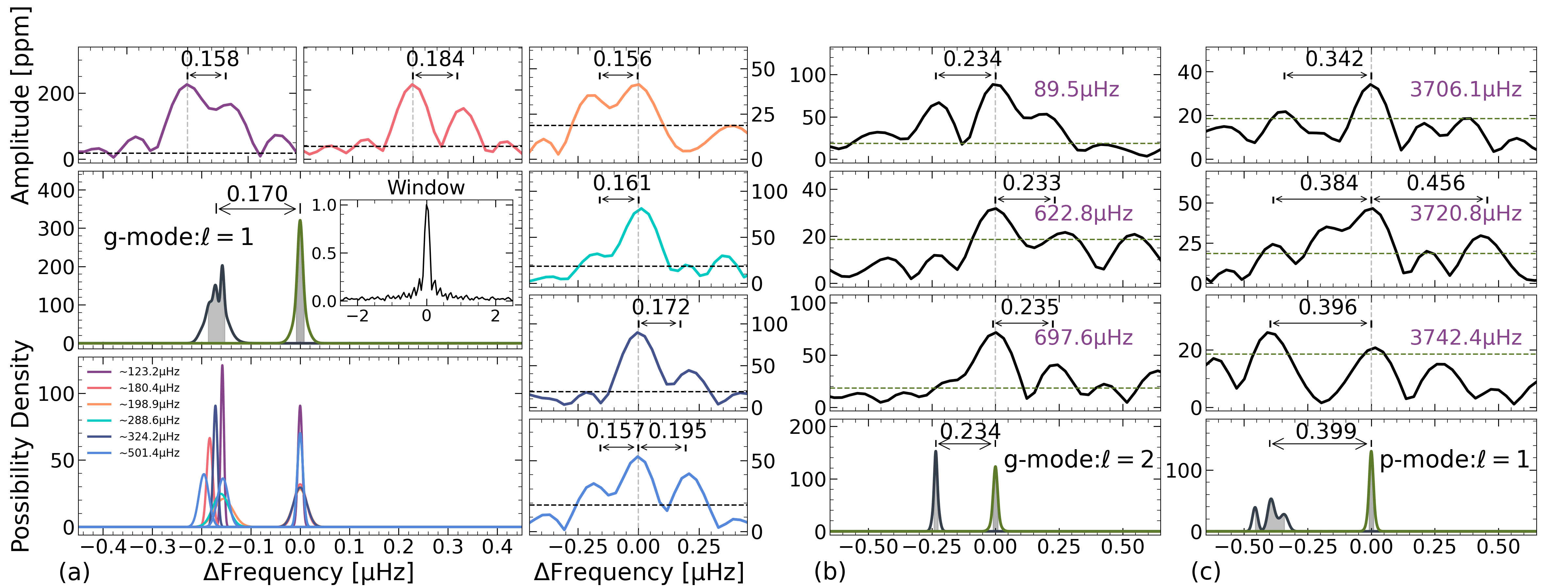}
    \caption{Likely multiplet frequencies induced by rotation detected in EPIC~220422705. (a) Six {\sl g}-mode triplet components. The  multiplets are shown in the top and right panels, with the horizontal lines indicating the 5.2$\sigma$ threshold and the frequency spacings are given in each panel. Different frequencies are marked by their colors provided in the bottom-left panel where each Gaussian distribution represents the probability density of a component by their shifted frequency and error. In the middle-left panel, the superposition function of all components gives the probability density of frequency spacing and the area in shadow defines the probability of 68.27\% that is identical to 1$\sigma$ of $N\sim(0,1)$. (b) Similar to (a) but for three incomplete {\sl g}-mode quintuplets. (c) Similar to (a) but for three {\sl p}-mode triplet components.}
    \label{fig:fre_splitting}
\end{figure*}

From linear perturbation theory, an eigenmode of oscillation can be characterized by spherical harmonics that are described by three quantum numbers: the radial order $n$, the degree $\ell$, and the azimuthal order $m$.
When a star rotates, the degenerated $m$ components will split into $2\ell+1$ multiplets.
Referring to \cite{1951ApJ...114..373L}, their frequencies are related by,
\begin{equation}
    \label{eq:fre_splitting}
    \nu_{n,l,m} = \nu_{n,l,0} + m\Omega(1-C_{n,\ell}),
\end{equation}
where $\nu_{n,l,0}$ is the frequency of the central $m=0$ component, $\Omega$ is the solid rotational frequency, and $C_{n,\ell}$ is the Ledoux constant. For acoustic {\sl p}-mode, $C_{n,\ell}$ is very near to zero and can be ignored, whereas it is estimated as $C_{n,\ell} \sim 1/ \ell(\ell +1)$ for high-radial order gravity {\sl g}-modes.

To resolve any rotational split multiplet  from spaceborne photometry, a minimum criterion is that the observations should cover at least  twice the rotation periods. \citet{2018OAst...27..112C} and \citet{2021MNRAS.500.2461S} present the distribution of rotation periods for sdB stars determined from {\em Kepler} photometry. Frequency multiplets found rotation periods from a few days to near one year with most having periods a bit longer than one month. This indicates that {\em K}2 photometry can likely resolve frequency multiplets in sdB stars. 

Figure\,\ref{fig:fre_splitting} shows the frequency spacings of 12 groups of frequencies. We first consider six {\sl g}-mode frequency groups that are detected with close frequency spacings around 0.17~$\mu$\,Hz, which we consider to be dipole modes, i.e., $f_5\sim$ 123.09~$\mu$Hz, $f_{14}\sim$ 180.39~$\mu$Hz, $f_{41}\sim$ 198.89~$\mu$Hz, $f_{23}\sim$ 288.56~$\mu$Hz, $f_{19}\sim$ 324.16~$\mu$Hz and $f_{30}\sim$ 501.25~$\mu$Hz. The weighted (by $1/\sigma f_i$) average value is $0.168\pm 0.016~\mu$Hz. Those dipole modes give a rotational frequency of 0.33~$\mu$Hz which would mean quintuplet splitting of $0.28~\mu$Hz using $C_{n,1} = 1/2$ and $C_{n,2} = 1/6$. Three {\sl g}-modes, $f_{18}\sim$ 89.48~$\mu$Hz, $f_{54}\sim$ 622.79~$\mu$Hz and $f_{24}\sim$ 697.63~$\mu$Hz, have close frequency spacings of $\sim 0.23$~$\mu$Hz, which could be rotational quintuplets, considering frequency uncertainties. We also resolve three $\ell=1$ {\sl p}-mode multiplets, $f_{52}\sim$ 3706.12~$\mu$Hz, $f_{39}\sim$3720.77~$\mu$Hz and $f_{62}\sim$3741.96~$\mu$Hz, with low-amplitude peaks at a frequency distance of $\sim0.4~\mu$Hz. 

In order to determine the rotational period in a quantitative way, we propose a new approach that defines the rotation period associated with errors by their probability of occurrence. As shown in  Figure\,\ref{fig:fre_splitting}, we adopt a Gaussian distributions, $N \sim (\delta f_i, \sigma f_i^2)$, to represent the probability of a group of resolved frequencies at their shifted values. Here $\delta f_i $ is the relative frequency to the central component. The probability of 68.27\% (i.e., 1$\sigma$ in $N\sim(0,1)$) was calculated to define the values and the uncertainties of frequency spacings. We obtained the values of  $0.170\pm0.049$~$\mu$Hz, $0.234\pm0.035$~$\mu$Hz and $0.399\pm0.132$~$\mu$Hz for $\ell=1$, and $2$ {\sl g}-modes and {\sl p}-modes, respectively.
The corresponding rotation periods are $34.04_{-7.12}^{+13.78}$~d, $41.21_{-5.35}^{+7.22}$~d and $28.86_{-7.21}^{+14.34}$~d, respectively.

Our result suggests that a slightly differential rotation occurs in EPIC~220422705 as {\sl g}- and {\sl p}-modes probe stellar interior under different depth  \citep[see, e.g.,][]{2022arXiv220111629K}. We note that our results are completely based on only marginally-resolved frequencies with low-amplitudes near the detection limit. We do not detect multiplets in the four highest-amplitude frequencies or 12 of the 13 highest-amplitude frequencies. Nevertheless, the rotation period we determine is consistent with that of a typical sdB star \citep[see details in][]{2018OAst...27..112C,2021MNRAS.500.2461S}. \citet{2015ApJ...805...94F} claims to detect a differential rotation in KIC~3527751 whose core rotates slower than the envelope, which, however, was challenged by an independent analysis of the same photometry by \citet{2018ApJ...853...98Z}, citing that the claimed rotational {\sl p}-mode multiplets  had missing components under significant confidence. In reverse, \citet{2005ApJ...621..432K} suggest that the core might rotate faster than the envelope of sdB stars from evolutionary models. In the case of EPIC~220422705, a binary system but with a poorly measured orbital period \citep{2012ApJ...758...58B}, a slightly faster rotating envelope could be interpreted that the orbital companion has accelerated it via the tidal force. Theory predicts that angular momentum transportation leads to radiative envelope first synchronized then gradually proceeds to the inner part \citep{1989ApJ...342.1079G}. In combination with the poorly-determined orbital and rotation period (because of low-amplitude multiplets), it would be unwise to speculate too much on this star. There are other sdBV stars which would be better for such work. To be cautious, EPIC~220422705 can still be a rigid object if the uncertainties of  rotational periods are fully considered.

\subsection{Period spacing}\label{sec:Period_Spacing}
\begin{figure*}
    \centering
    \includegraphics[width=\textwidth]{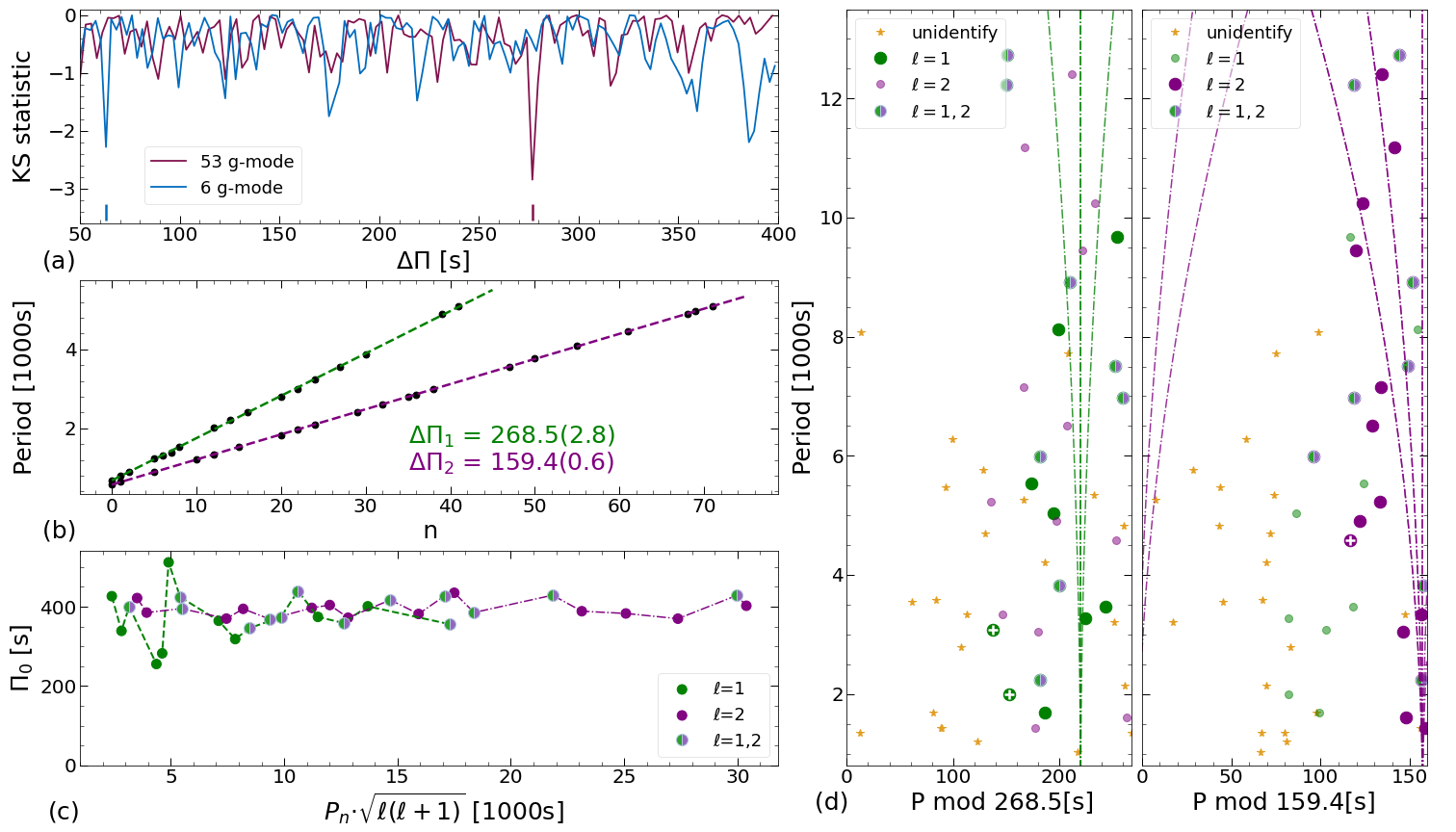}
    \caption{Period spacing and mode identification for independent {\sl g}-modes. (a) Kolmogorov-Smirnov (KS) tests on two period sets. The vertical segments locate the minimum values for preliminary spacing.
    (b) The linear fitting for the identified  $\ell = 1$ and $\ell = 2 $ modes, respectively, where the slopes indicate the values (text) of their period spacing.
    (c) The reduced period spacing, $\Pi_0 = \sqrt{\ell(\ell + 1)} \cdot \Delta \Pi_{\ell}$, as a function of the reduced pulsation period.
    (d) \'Echelle diagram for dipole (left) and quadruple (right) modes. The vertical curve locates the corresponding period for rotational components, 3 and 5 for triplet and quintuplet, respectively.
    The modes marked with white '+' get larger period deviations }
    \label{fig:PS}
\end{figure*} 
For {\sl g}-mode pulsations in the asymptotic regime, consecutive high-radial orders ($n \gg \ell$) follow a pattern of equal period spacing \cite[see, e.g., ][]{2010aste.book.....A}, which depends on the structure. Seismology theory provides the following relationship,  
\begin{equation}\label{eq1}
	\Delta \Pi_{l}\approx  \frac{\Pi_{0}}{\sqrt{(l(l+1))}}, 
\end{equation}
with $\Pi_{0}$ defined as,
\begin{equation}\label{eq2}
     \Pi_{0} = 2 \pi^2(\int_{1}^{R} \frac{N}{r}dr)^{-1},
\end{equation}
where $N$ is the Brunt-V\"ais\"al\"a~frequency and $r$ is the radial coordinate.
For the period spacing of $\ell$ = 2 sequence, it is related to the $\ell$ = 1 sequence as,
\begin{equation}
    \label{eq3}
    \Delta\,\Pi_{\ell=2} = \frac{1}{\sqrt{3}} \Delta\Pi_{\ell=1}.
\end{equation}

Previous analysis of sdBV stars from {\em Kepler} and TESS reveals that the period spacing is about 250~s and 150~s for dipole and quadruple modes, respectively  \citep[see, e.g.,][]{2011MNRAS.414.2885R,2020MNRAS.495.2844S}. To find the spacing periods in EPIC~220422705, we performed the popular Kolmogorov-Smirnov (KS) test on the independent {\sl g}-mode frequencies. The KS test returns spacing correlations as highly-negative values for the most common spacings in a dataset \citep{1988IAUS..123..329K}. We first apply the KS test to the six rotational (incomplete) triplets, which has a deepest trough at 63~s = 252/4  (Figure\,\ref{fig:PS} a). In addition, we perform a linear fitting to those 6 periods with a result of $\sim$265.5~s. Then we applied another KS test for the 53 independent frequencies lower than 1000~$\mu$Hz as probable {\sl g}-modes, which gives a value around 276.8~s for dipole modes (Figure\,\ref{fig:PS}\,a). All values are consistent with that of dipole modes in sdB stars  \citep{ 2011MNRAS.414.2885R,2020MNRAS.495.2844S}. 

Based on the preliminary period spacings, we have identified nine modes as dipole, 13 as quadruple, and additional 8 frequencies which fitted both period sequences. We note that the four identified modes include one of the above 6 dipole modes, $f_{5}\sim 123.1~\mu$Hz. We obtained the average period spacing of 268.5 $\pm$ 2.8~s and 159.4 $\pm$ 0.6~s for $\ell = 1$ and $\ell = 2$ modes, respectively, via linear fitting to 17 (4 + 5 + 8) dipole modes and 21 (13 + 8) quadruple modes (Figure\,\ref{fig:PS}\,b). We list $\ell$ and relative $n$ values in Table\,\ref{Tab:freq}. We note that the real radial order can only be obtained through seismic modeling. Our results suggest that EPIC~220422705 has a somewhat large period spacing among the known sdB variables, in a range of [220, 270]~s for the dipole mode \citep[see, e.g.,][]{2011MNRAS.414.2885R,2020MNRAS.495.2844S,2021A&A...651A.121U}. As stellar models presented in \citet{2021A&A...651A.121U}, the evolution tracks suggest that the lower value of period spacing, the lower value of $\log g$. This agrees well to a low $\log g$ derived for EPIC~220422705. However, atmospheric parameters, with a much higher precision, are encouraged for EPIC~220422705 to test the results of \citet{2021A&A...651A.121U} in future. 

The {\'echelle} diagrams for two sequences are presented in Figure\,\ref{fig:PS}\,(d) where the $\ell =2$ sequence is more consistent than the $\ell =1$ sequence. There are two $\ell =1$ frequencies with larger period deviations, $f_{19} \sim 324.16\,\mu$Hz, $f_{30} \sim 501.25\,\mu$Hz whereas only one occurs in the $\ell =2$ sequence, $f_{8} \sim 218.36\,\mu$Hz. 
Asymptotic theory indicates that period spacings are determined by the size of the pulsation resonant cavity \citep[see e.g.][]{1980ApJS...43..469T} and could be affected by the extent of the convective core \citep{2007A&A...465..509S}. The ideal pattern for a period sequence in the \'echelle diagram is a vertical ridge for central components of a star with internally-homogeneous composition. 
Some deviations from the mean period spacing are to be expected in {\sl g}-mode pulsating sdB stars -- and indeed have already been unambiguously detected in some cases \citep{2014A&A...569A..15O} -- due to the phenomenon of mode trapping by steep composition gradients. Such deviations reflect properties of the stellar interior, and in particular could reveal the character of the mixing processes at work near the core boundary \citep[see, e.g.,][]{2011A&A...530A...3C,2017MNRAS.465.1518G}. Figure\,\ref{fig:PS}\,(c) shows the deviation of period spacing as a function of the reduced pulsation period. We only observe a large deviation that might be associated to a trapped mode at the fifth order. Indeed, seismic models suggest that strong trapping are more likely found for the lower order (higher frequency or shorter period) modes than the higher order {\sl g}-modes \citep{2014ASPC..481..179C}.

\section{Amplitude and Frequency Modulations}\label{sec:AF}
This section provides our methodology and characterization of amplitude and frequency modulations (AM/FM) for the most significant frequencies. In practice, we follow the processes as described in \citet{2018ApJ...853...98Z} to extract frequency information of subsets of the entire light curve. Here the time interval and window width are 1~d and 30~d, respectively. If close peaks within the frequency resolution ( $\sim$ 0.4 $\mu$Hz) are detected, we keep the highest peak as the measured value for that frequency. In order to measure AF/FM significantly, frequencies should have amplitudes above 8.8$\sigma$ of the local noise in each piece of the light curve. This ultimately leads to 22 frequencies that could be analyzed for AMs/FMs, which are marked in the last column of Table~\ref{Tab:freq}.

\subsection{The fitting method}
\begin{figure*}
    \centering
    \includegraphics[width=0.48\textwidth]{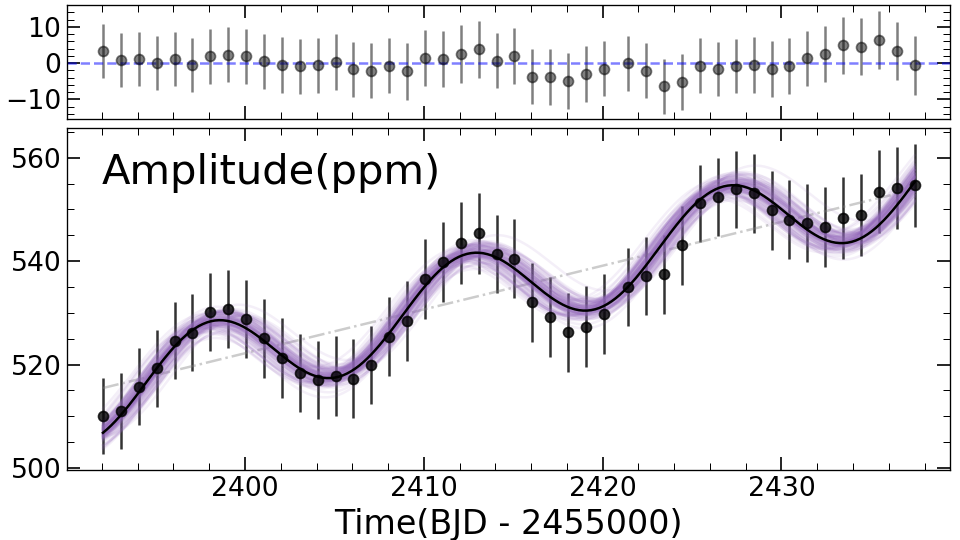}
    \includegraphics[width=0.48\textwidth]{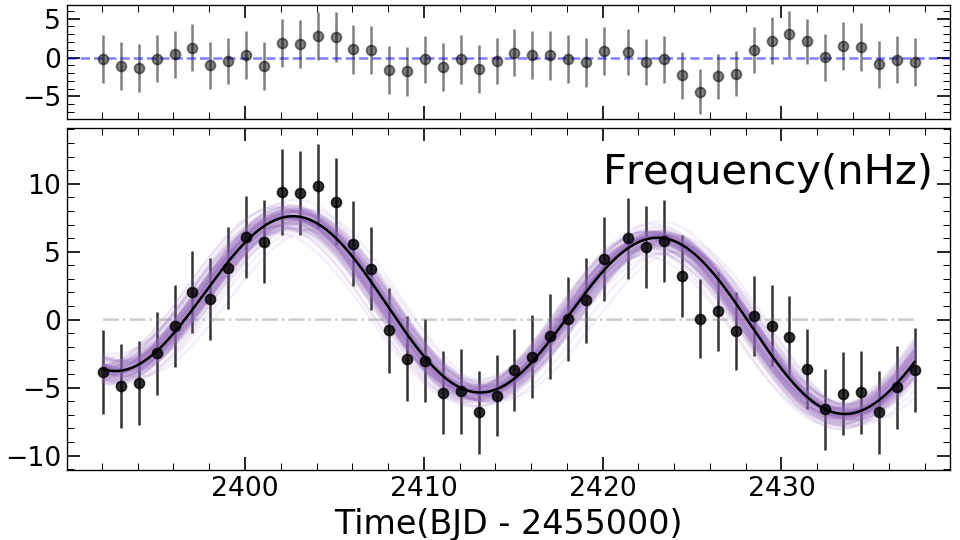}
    \includegraphics[width=0.48\textwidth]{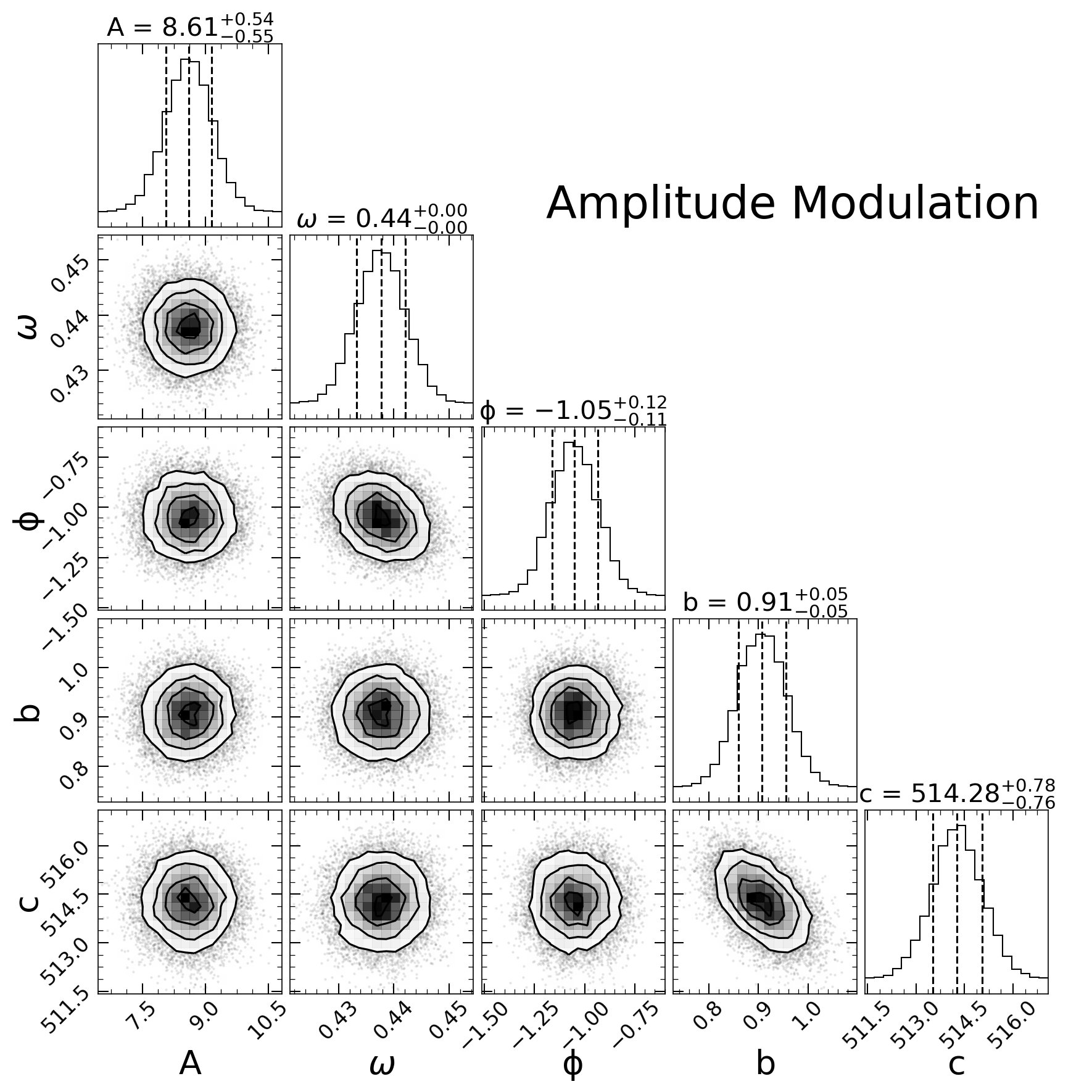}
    \includegraphics[width=0.48\textwidth]{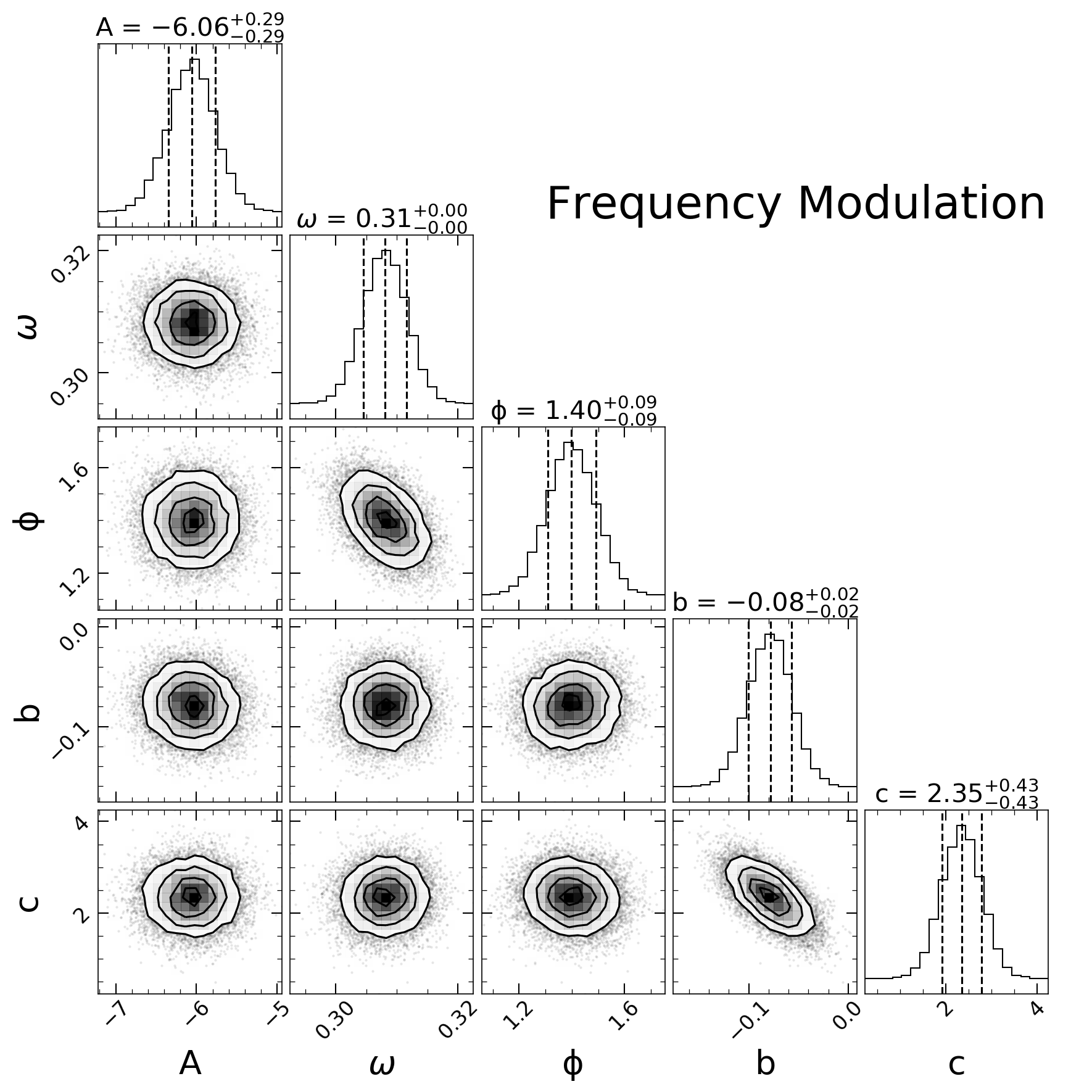}
    \caption{A demonstration of amplitude and frequency modulation for the frequency $f_{3}\sim$328.75\,$\mu$Hz. 
    The top panels show precise patterns of AM and FM together with the residuals after the best fittings, indicated by solid curves. The shadow curves indicate 100 confidence fitting curves that are randomly taken from \texttt{MCMC} chains. The dashed lines indicate linear relationships to the modulated patterns or constant to zero. The bottom panels show the distributions and the correlations of the five fitting parameters by the \texttt{MCMC} method. The vertical lines, from left to right, define the confidential fittings and stand for 16, 50 and 84 percentiles. The contours are at the 1$\sigma$, 2$\sigma$ and 3$\sigma$ levels, respectively.
    }
    \label{fig:corner}
\end{figure*}

A quick look of all AM/FM patterns occurring in these 22 frequencies suggest that most of them exhibit simple or quasi-regular variations (Figures~\ref{fig:corner}, \ref{fig:AF}, \ref{fig:AF_R} and \ref{fig:697}). In order to quantitatively characterize the modulation patterns, we apply three simple types of fittings: linear, parabolic, and sinusoidal waves. We first calculate standard deviations of our sub-set data, $\sigma_0$. Then we adopt a simple AM/RM fitting, typically linear first and then a second type on the residuals. This fitting method may be iterated several times until the residuals present no clear structure and look like random noise. For each fitting, a standard deviation of residuals will be calculated as:
\begin{equation}\label{eq:sigma_jk}
	\sigma_{k}^2 = \frac{1}{N} \sum_{i=1}^{N}[m_i - G_{k}(t_i)]^2,
\end{equation}
where $G_{k}(t)$ defines as,
\begin{equation}
\begin{aligned}
&G_{k}(t)=\left[g_1(t),g_2(t),g_3(t),g_1(t)+g_3(t),g_2(t)+g_3(t)\right],\\
&\left\{
\begin{array}{l}
    g_1(t) = bt+c,\\
    g_2(t) = at^2+bt+c,\\
    g_3(t) = A\sin(\omega t + \phi)+A_0,
\end{array} 
\right.
\end{aligned}    
\end{equation}
Here $k = 1, 2, 3, 4, 5$ and $N$ denotes number of data points, $m_i$ are the measured values of amplitude or frequency.

In principle, the freedom degree of $G_k(t)$ increases as $k$ increases, which in turn results in a lower $\sigma_{k}$. However, to avoid overfitting of the modulation patterns, for instance, by including a linear or parabolic fitting, we follow the statistical test by \cite{1975MNRAS.170..633P} who defines $\lambda$ as,
\begin{equation}
	\lambda = \dfrac{(\sigma_{1}^2 - \sigma_{2}^2)/(D_2-D_1)}{\sigma_{2}^2/(N-D_2)}.
\end{equation} 
Here $D_i$ is the freedom degree of the fitting function, e.g., 2 and 3 for linear and parabola fitting, respectively. A significant improvement for the higher freedom degree fitting should have the parameter that meets the $F-$distribution, $\lambda > F_{P=99.75\%}(D_2-D_1,N-D_2)$. For each AM/FM, we prefer to keep the fitting function with a lower freedom degree as indicated by the parameter $\lambda$. The results of our fitting for each AM/FM is listed in Table~\ref{Tab:AFM_D} where most frequencies have a sinusoidal component, $G_3(t)$, or with an additional linear fitting, $G_4(t)$.

After the fitting function was set, we adopted the posterior distributions based on the Bayesian frame to estimate the best fitting parameters and their uncertainties. The posterior distributions of parameters are sampled by the Markov Chain Monte Carlo (\texttt{MCMC}), which is implemented by the \texttt{EMCEE} code \citep{2013PASP..125..306F}. 
The \texttt{MCMC} sampling is performed with more than 22 chains, according to the number of free parameters. It proceeds until the chains converge to values that are inferred by the auto-correlation time module of the \texttt{EMCEE}. Parameters of the best fittings are given by the medians of marginalized posterior distributions, associated with the corresponding errors that are calculated by the half-widths between the 16$^{\rm th}$ and 84$^{\rm th}$ percentiles of the distributions. 

Figure \ref{fig:corner} is an example of the \texttt{MCMC} application for AM and FM occurring in the frequency $f_{3} \sim 328.75~\mu$Hz. We clearly see that both AM and FM exhibit periodic variations but with different periods and phases. In general, we find that almost all measurements are consistent with the fitting curves, accounting for uncertainties. The best fittings return $A = 8.61\pm0.54$\,ppm and $6.1\pm0.3$\,nHz, $T = 14.4\pm0.1$\,d and $20.4\pm0.2$\,d, $\phi = 5.2\pm0.1$ and $1.4\pm0.1$, $b = -0.91\pm0.05$ and $-0.08\pm0.02$, and $c = 514\pm1$ and $2.4\pm0.4$ for AM and FM, respectively.

\subsection{Characterization of modulation patterns}

\startlongtable
\begin{deluxetable*}{lcrcrrrrrrrc}
    \label{Tab:AFM_D}
    \input{AFM_Details}
\end{deluxetable*}

\begin{figure*}
    \centering
    \includegraphics[width=0.2693\textwidth]{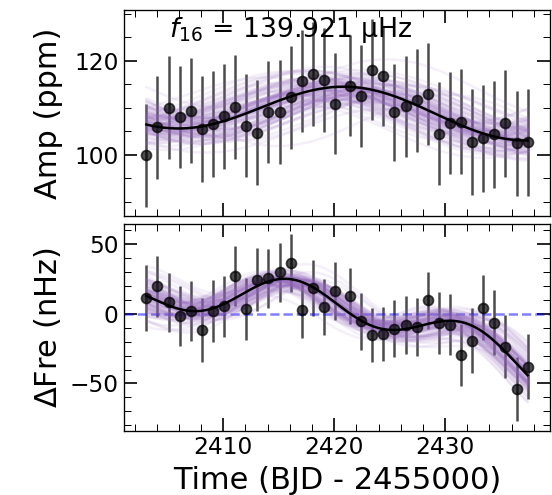}
    \includegraphics[width=0.234\textwidth]{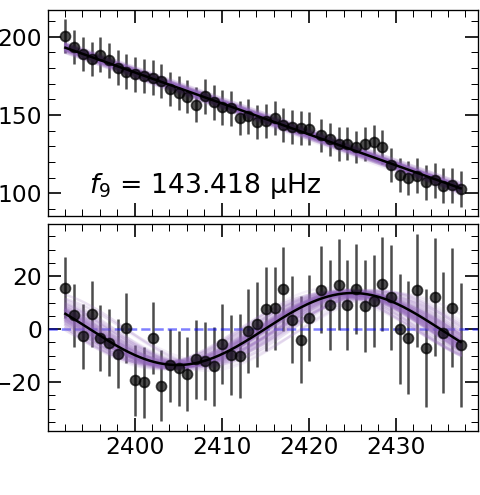}
    \includegraphics[width=0.234\textwidth]{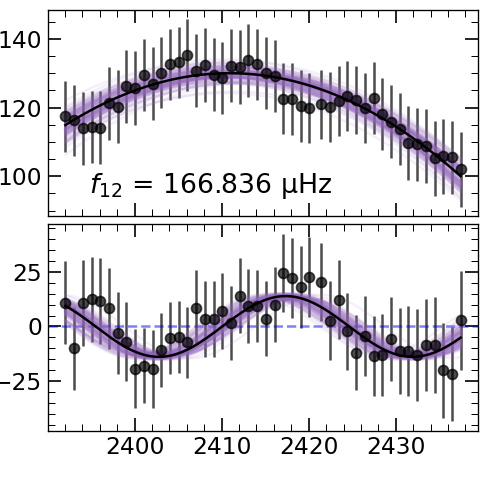}
    \includegraphics[width=0.234\textwidth]{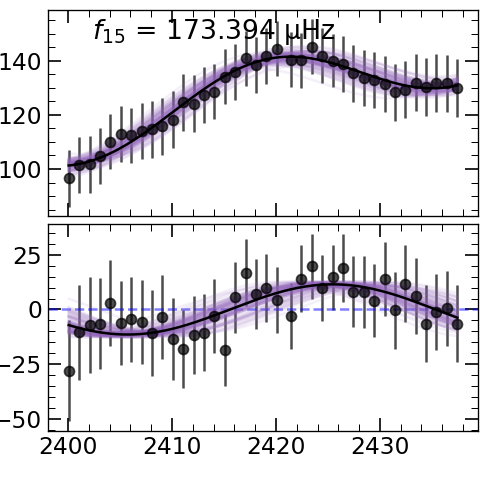}\\
    \includegraphics[width=0.2693\textwidth]{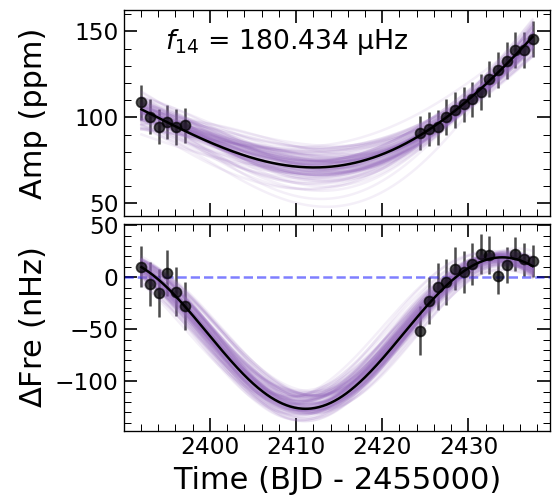}
    \includegraphics[width=0.234\textwidth]{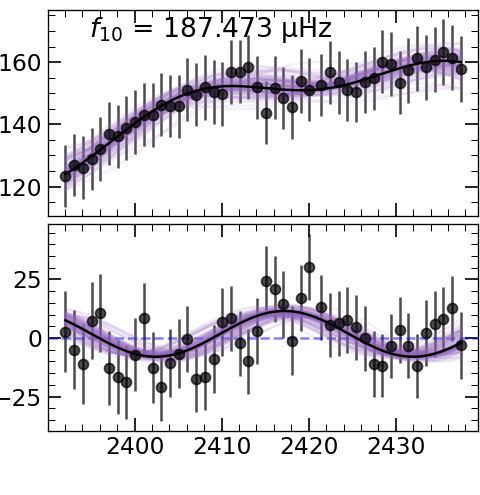}
    \includegraphics[width=0.234\textwidth]{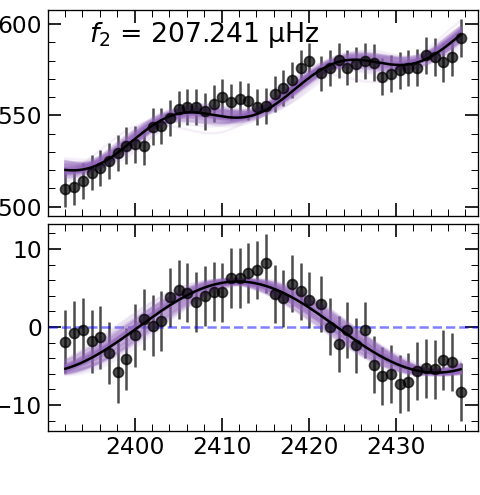}
    \includegraphics[width=0.234\textwidth]{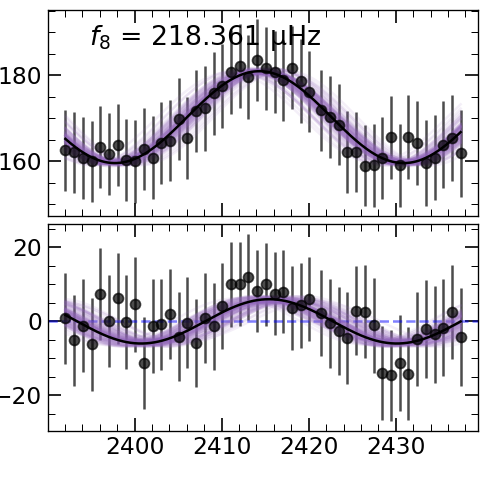}\\
    \includegraphics[width=0.2693\textwidth]{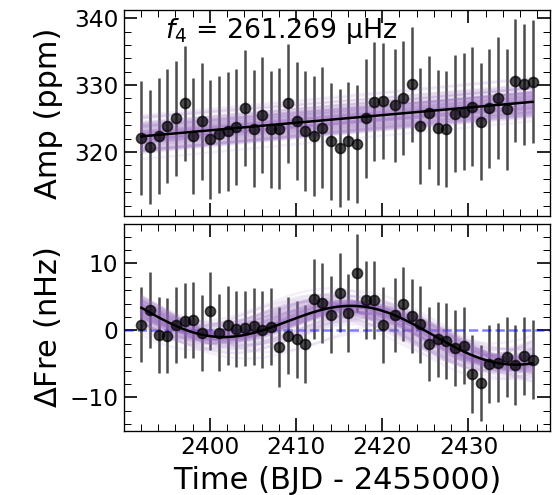}
    \includegraphics[width=0.234\textwidth]{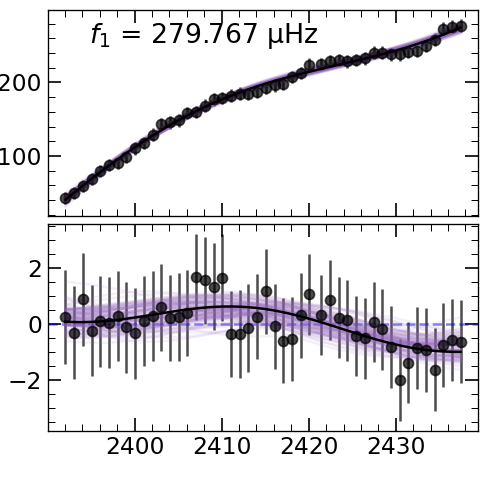}
    \includegraphics[width=0.234\textwidth]{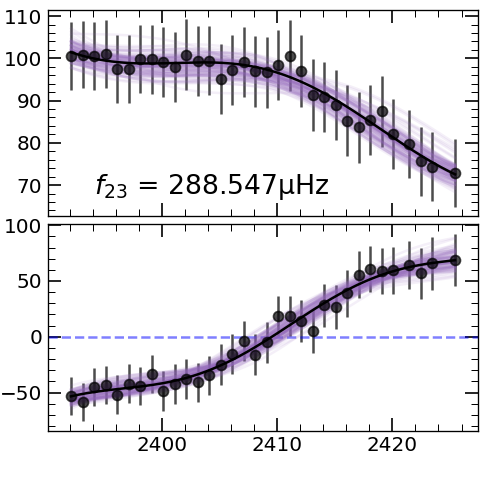}
    \includegraphics[width=0.234\textwidth]{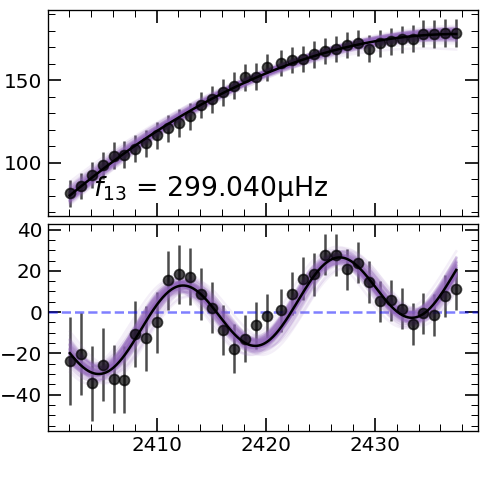}\\
    \includegraphics[width=0.2693\textwidth]{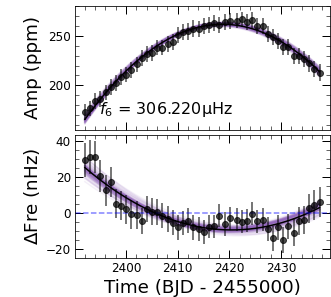}
    \includegraphics[width=0.234\textwidth]{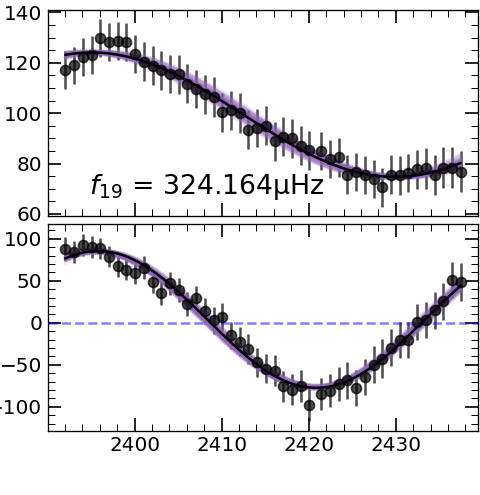}
    \includegraphics[width=0.234\textwidth]{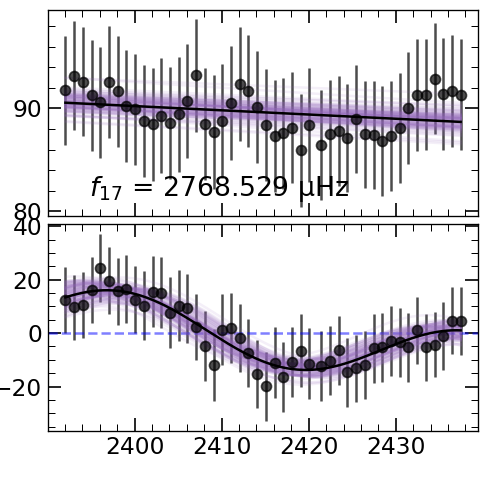}
    \includegraphics[width=0.234\textwidth]{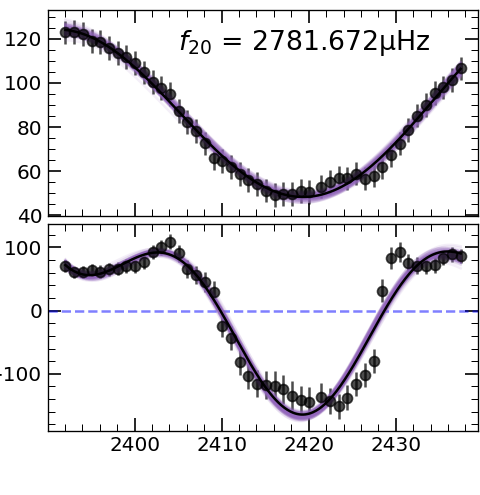}
    \caption{A gallery of 16 frequencies with AM/FM variations. Each module refers to the AM (top panel) and FM (bottom panel) variations of one single frequency. The frequencies are shifted to the average values as represented by the dashed horizontal lines, which are indicated in the bottom panel. The solid curves in purple and black represent the fitting results from the {\texttt MCMC} method and the optimal fitting, respectively.
    }
    \label{fig:AF}
\end{figure*}

\begin{figure*}
    \centering
    \includegraphics[width=0.2693\textwidth]{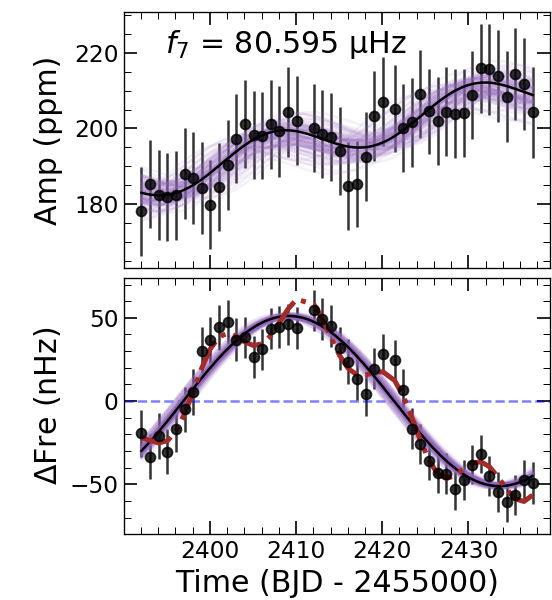}
    \includegraphics[width=0.234\textwidth]{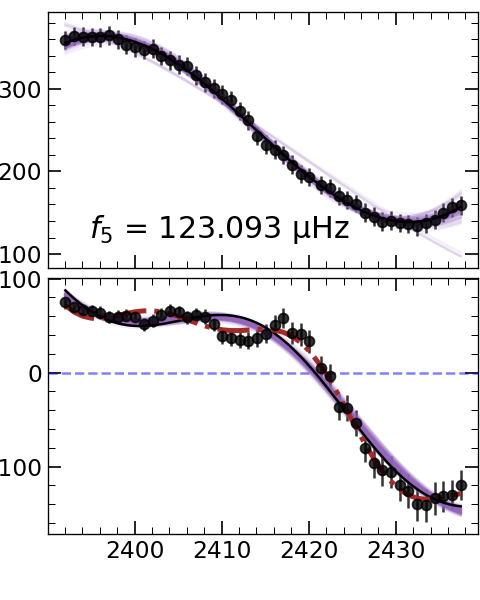}
    \includegraphics[width=0.234\textwidth]{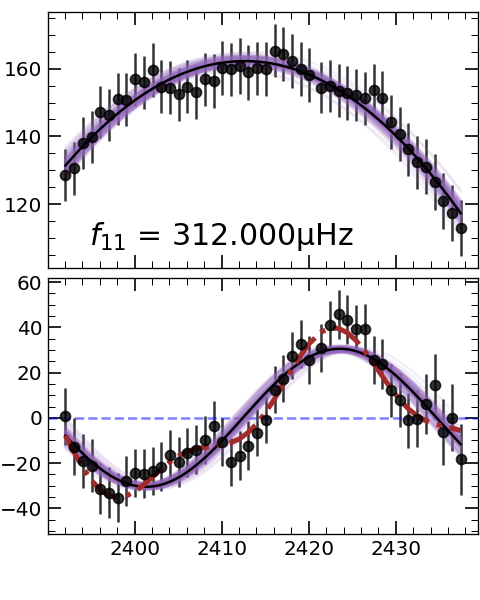}
    \includegraphics[width=0.234\textwidth]{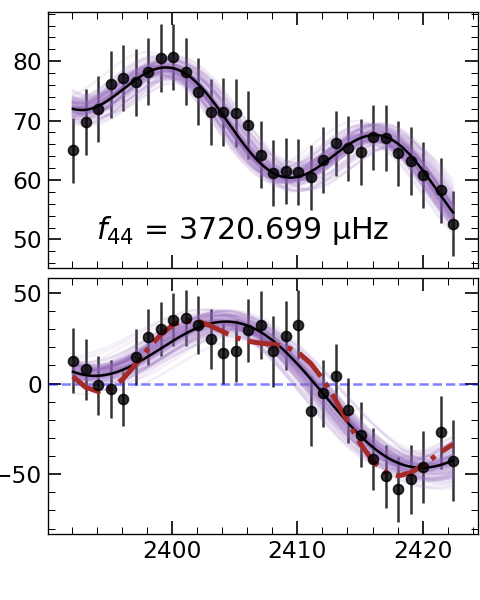}
    \caption{Similar to Figure\,\ref{fig:AF} but for four frequency residuals still showing regular patterns. The dashed (red) curves refer to the final fitting with additional components together with the $G_{k}$ function.}
    \label{fig:AF_R}
\end{figure*}

In this section, we describe the modulation patterns in AM/FM of all 22 significant frequencies except the frequency of $f_{3}$ which has already been described. Figure\,\ref{fig:AF} is a gallery of 16 frequencies with various modulation patterns, which are identified by different fitting function $G_k(t)$ and summarized in Table~\ref{Tab:AFM_D}. In general, the fitting periods of AMs and FMs are on timescale of months. 

We now specifically describe these AMs and FMs. A few modes have been observed with simple modulation patterns of only linearly decreasing or increasing in amplitude or frequency. For instance, the amplitude of $f_9 \sim 143.4~\mu$Hz decreases from 200~ppm to 100~ppm over the time interval of about 50~days. In other observed minor cases, a few modes can be characterized by a simple parabolic fitting, such as the AM/FM of $f_6 \sim 306.2~\mu$Hz where both the amplitude and frequency reach to their vertex values near the time BJD~=~2457420~d. The majority of modulations we observed have quasi-sinusoidal patterns, most with extra linear and a few with parabolic fittings. 
For example, the amplitude and frequency of $f_{8} \sim 218.4~\mu$Hz exhibit completely sinusoidal pattern and evolve in phase. Whereas the frequency of $f_{13} \sim 299.0~\mu$Hz demonstrates a sinusoidal pattern with an additional linear fitting. We only find $f_{16}\sim 139.9~\mu$Hz whose frequency and $f_{10}\sim 187.5~\mu$Hz whose amplitude show a sinusoidal plus a parabolic pattern. We note that the amplitude of $f_{14} \sim 180.434~\mu$Hz decreases below the detection threshold from $\mathrm{BJD}-24554000 = 200$~d to 225~d, which holds even if our subsets span 70\,d. For most frequencies, a variation of frequency scale is around 20~nHz but there are a few exceptions with large values around 100~nHz. For instance, $f_{20}\sim2781.672~\mu$Hz, $f_{19}\sim324.164~\mu$Hz and $f_{23}\sim288.547~\mu$Hz. The variations of amplitude can span up to a few hundred ppm ($f_{1}\sim279.767~\mu$Hz) or down to a few ppm ($f_{4}\sim261.269~\mu$Hz). We note that the amplitudes of AM and FM are not strictly proportional to each other. For instance, the amplitude of $f_1$ increases from about 50~ppm to 250~ppm, but the frequency only varies around 1~nHz. 

A very interesting feature of the observed AM/FM is that several frequencies are found with (anti-) correlations between AM and FM. We thus calculate the values of correlation for all frequencies as listed in Table~\ref{Tab:AFM_D}. The frequencies $f_{23}\sim288.561\mu\mathrm{Hz}$ and $f_{6}\sim306.223~\mu$Hz all exhibit strong correlation with a coefficient $|\rho_\mathrm{a,f}|>0.8$ between their amplitude and frequency variation. 
For instance, $f_{23}$ and $f_{6}$ show clear anti-correlation between AM and FM, with derived coefficients of -0.867 and -0.824, respectively. For $f_{23}$, the amplitude exhibits a slightly decreasing trend whereas the frequency is opposite. For $f_{6}$, its AM and FM are both represented by parabolic fits but with anti-phase evolutions. We also note several frequencies with correlated AM and FM patterns showing sinusoidal variations, such as  $f_{15} \sim 173.4~\mu$Hz and $f_{8} \sim 218.3~\mu$Hz.

Figure\,\ref{fig:AF_R} shows another 4 frequencies, $f_5\sim123.087~\mu\mathrm{Hz}$, $f_7\sim80.595~\mu\mathrm{Hz}$, $f_{11}\sim312.0~\mu\mathrm{Hz}$ and $f_{39}\sim3720.7~\mu\mathrm{Hz}$, whose FM patterns still present modulation structures after fitting by a $G_k(t)$ function with $k=4,5$. To remove those FM residuals, they needed additional sinusoidal function, i.e., three fitting functions to present FM patterns. 
For a strong correlation frequency, $f_5$, with $\rho_\mathrm{a,f} = 0.862$, exhibits regular AM and FM with period of $101^{+14}_{-14}$ and $40^{+1}_{-1}$ days, respectively, derived from \texttt{EMCEE}.
The other three frequencies, $f_7$, $f_{11}$ and $f_{39}$ are determined by \texttt{EMCEE} with either different fitting types of functions or different periods of sinusoidal patterns between AM and FM. For instance, the periods of AM and FM in $f_7$ are calculated as $23^{+1}_{-1}$ and $49^{+1}_{-1}$ days, respectively, almost with a ratio of $1:2$. The \texttt{EMCEE} returns periods with values of $\sim 10.3, 19.0, 17.4$ and $11.3$~days for the additional sinusoidal fittings of the FM residuals of $f_7$, $f_5$, $f_{11}$ and $f_{39}$ after extracting $G_k(t)$, respectively. We note that FM periods of these residuals are shorter than the periods fitted by $G_k(t)$.

\begin{figure}
    \centering
    \includegraphics[width=0.5\textwidth]{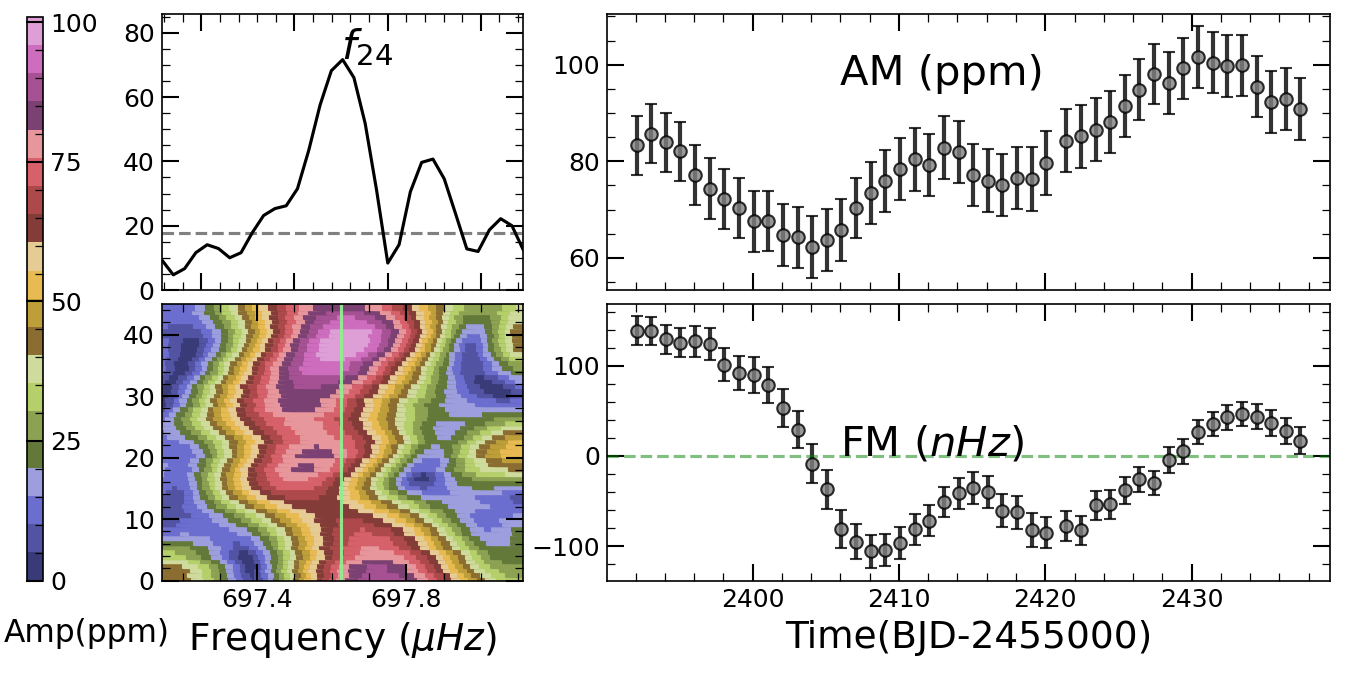}
    \caption{AM and FM of $f_{24} \sim 697.6\,\mu$Hz, a frequency that cannot be fitted by simple function of $G_{k}$. The top left panel shows the local LSP where the dashed horizontal line is the 5.2~$\sigma$ threshold. The bottom left panel provides the sLSP around that frequency and the color bar (indicating amplitude) is shifted to the leftmost side, in which the vertical line is the averaged frequency. Precise measurements of amplitude and frequency variations are presented in the top- and bottom-right panels, respectively. The frequencies are shifted to its average as indicated by the horizontal line. 
    }
    \label{fig:697}
\end{figure}

Figure~\ref{fig:697} presents the AM and FM of the frequency $f_{24}\sim697.62~\mu$Hz that shows a quasi-regular behavior but cannot be described by simple fittings of $G_k$ functions, which is also suggested by the fine profile of the LSP and the sliding LSP (sLSP). Thus we do not perform fitting and \texttt{EMCEE} on its AM and FM. 
We observe that both the AM and FM began with a decreasing trend: the amplitude went down from $\sim80$~ppm to a local minimum value of $\sim60$~ppm and the frequency varied from +120\,nHz down to -100\,nHz relative to its averaged frequency. Then the frequency and amplitude had experienced an increasing trend. The amplitude reaches to its maximum with a time interval of about 30 days but passing across one stationary point, whereas the frequency generally went up to around the average value with a back and forth trend.

\section{Discussion}
\label{discuss}

\begin{figure}
    \centering
    \includegraphics[width=0.49\textwidth]{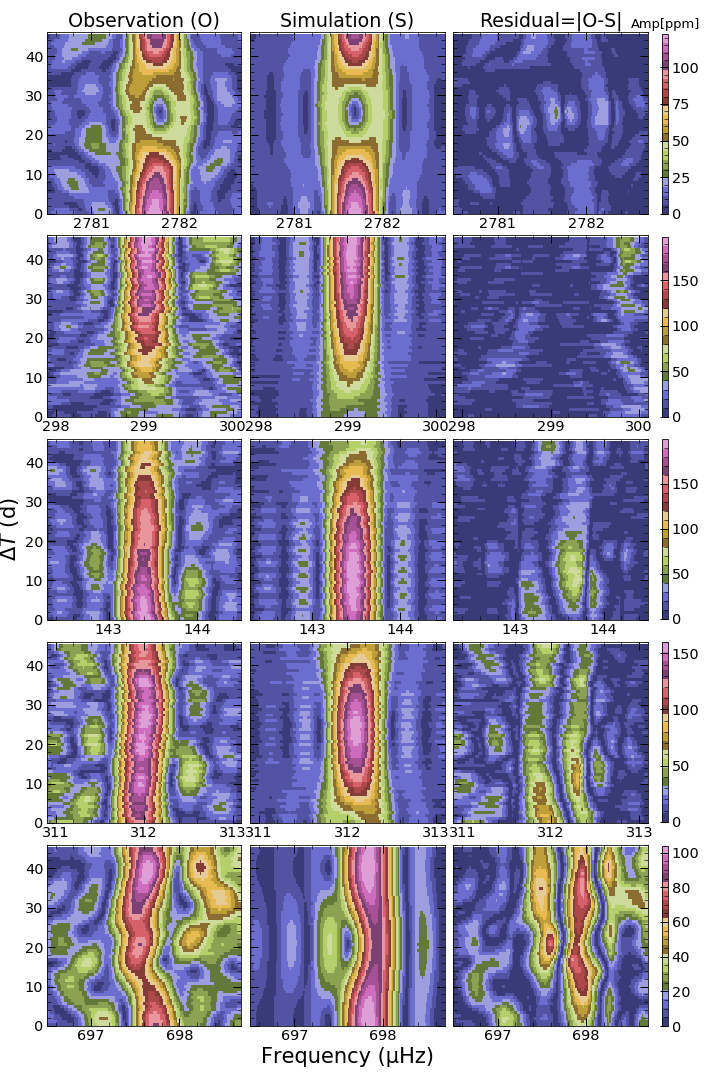}
    \caption{Comparison of sliding LSP of five representative frequencies between observations and simulations. The left, middle and right panels are the observed modulations, the simulated modulations and the residuals. The flatness of residuals fluctuate gradually worse from top to bottom panels. The residuals present double ridge structures if the sLSP cannot be simulated merely by two close signals.}
    \label{fig:OS}
\end{figure}

In this section, we will discuss the potential  interpretation of the observed AM and FM for all 22 frequencies of EPIC~220422705. Most of those modulations can be fitted with simple functions, $G_k(t)$, with most having sinusoidal fittings. They could be induced by the resonant coupling mechanism that predicts periodic amplitude and frequency modulation as a consequence of nonlinear weak interaction between different coupling modes \citep[see, e.g.,][]{1982AcA....32..147D,1995A&A...296..405B}.
However, the relative short duration of {\em K}2 photometry may suffer from other effects on the observed AMs and FMs, such as beating between unresolved frequencies. We therefore generate a series of quantitative simulations of close signals and compare the modulation patterns of those unresolved frequencies with the observed patterns of AM/FM.

\subsection{Presence of closely-spaced frequencies}\label{close_fre}
As presented in Section~2.3, we detected several potential rotation multiplets with frequency spacings of about 0.2 to 0.4\,$\mu$Hz, which is comparable to the frequency resolution determined by the observation duration, $\Delta f\approx 1/T$. This finding suggests that close frequencies, not only the multiplets we found, may be present in EPIC~220422705 with very high probability. In literature, many sdBV stars are resolved with such nearby frequencies from the original 4-yr {\em Kepler} observations \citep[see .e.g,][]{2018ApJ...853...98Z,2015ApJ...805...94F,2012MNRAS.424.2686B}. Those close frequencies will induce amplitude variations not frequency variations, called beating, if they are unresolved, as demonstrated in \citet{2018ApJ...853...98Z}. Here we only take two close frequencies as an example,
\begin{equation}\label{cf2}
z = A_1 \sin [2\pi (\omega_0 t + \phi_1)] + A_2 \sin [2\pi (\omega_0 t + \Delta \omega t + \phi_2)]. 
\end{equation}
If we consider two comparable amplitudes, $A_1 \sim A_2 = A$, and a very small frequency separation, $\Delta \omega \ll \omega_0$, the equation~(\ref{cf2}) will be reduced to,
\begin{equation}
\begin{aligned}
z &= 2A \cos(\pi \Delta \omega t + \phi{'}) \cdot \sin [2\pi (\omega_0 t + \Delta \omega t/2 + \phi)]\\
&\approx A(t)\cdot \sin(\omega_0t+\phi).
\end{aligned}
\end{equation}
The above equation indicates that two close frequencies will generate amplitude modulations if they were not resolved. We thus have to evaluate this effect on the observed AMs in EPIC~220422705.

A series of simulations had been performed to quantitatively compare the observed and calculated AMs. The simulation process are similar to that of \citet{2018ApJ...853...98Z}, including light curve construction and frequency injection, but adopted to {\em K}2 observations. In practice, we set three parameters $A_1$, $A_2$ and $\Delta\omega$
to be constant in  construction of each light curve based on equation~(\ref{cf2}) and phase to be zero for simplicity. We then make all parameters as variants in a series of 1331 light curves, with $A_1, A_2 \in [0.5, 0.6, ..., 1.5]\times A_0$ and $\Delta\omega \in [0.1, 0.12, 0.14, ... 0.3]~\mu$Hz. Then we transform the simulated light curves into sLSP and directly compare them with the observed sLSP via direct subtraction. Both types of sLSPs have to be transformed into the same time step and window length and scaled to the same maximum amplitude. The flatness of the residual sLSP defines the goodness of similarity between the two sLSPs. We finally select an optimal sLSP for the simulation and return the parameters.

Figure\,\ref{fig:OS} shows the comparison results for five representative frequencies, $f_{9} \sim 143.4\,\mu$Hz, $f_{13} \sim 299.0\,\mu$Hz, $f_{11} \sim 312.0~\mu$Hz, $f_{24} \sim 697.6~\mu$Hz and $f_{20} \sim 2781.6~\mu$Hz. We can conclude that the modulations of $f_{13}$ and $f_{20}$ are potentially induced by two close frequencies as revealed by no significant residuals. The modulation of $f_{11}$ and $f_{24}$ can hardly be simulated by two close frequencies, which indicates the AM and FM to be intrinsic, as revealed by the complex structure of the residual sLSP. The sLSPs of $f_9$ suggest that it is dominated by the beating effect but also experienced intrinsic AM and FM. We list the results of simulations in the last column in Table\,\ref{Tab:AFM_D} as: 'B' for the  completely beating effect and 'I' for intrinsic modulation.  Here we note that the index 'I' may also contain beating effect such as the close multiplets discovered but their AMs cannot completely represented by those beating effect. In summary, we found that five frequency modulations can be attributed to beating, $f_{5} \sim 123.1\,\mu$Hz, $f_{12} \sim 166.8\,\mu$Hz, $f_{2} \sim 207.2\,\mu$Hz, $f_{13} \sim 299.0\,\mu$Hz and $f_{20} \sim 2781.1\,\mu$Hz. Their modulations will not be discussed later. However, the other 17 frequencies are not well-represented by beating and must suffer from intrinsic modulations in amplitude and frequency.

\subsection{Potential interpretation of intrinsic modulation}
As stated above, in EPIC~220422705 many frequencies exhibit intrinsic AM and FM even though they may be contaminated by unresolved frequencies limited by {\em K}2 observing duration. These kinds of variations have previously been investigated for several sdBV stars \citep{2016A&A...594A..46Z,2018ApJ...853...98Z,2021ApJ...921...37Z}, showing several characteristics of their modulation patterns: stable, regular, irregular or complex features. In contrast to {\em Kepler} sdBV stars, the observed AMs and FMs here can only be determined for shorter temporal modulations or suffering from the beating of close frequencies. For instance, \citet{2016A&A...594A..46Z} found that in KIC~10139564 the modulating period is about 600~days for the dominant triplet which is much longer than the derived periods of months in EPIC~220422705.

From a theoretical perspective, we expect various types of AMs and FMs when perturbation theory is extended to nonlinear orders where different resonant modes can have weak interactions governed by amplitude equations \citep[see, e.g.,][]{1982AcA....32..147D,1984ApJ...279..394B,1995A&A...296..405B,1985AcA....35..229M,1997A&A...321..159B}. Both in multiplet resonance and direct parent-daughter resonance (e.g., $f_i + f_j \sim f_k$) the frequency mismatch, $\delta f = f_i + f_j - f_k$, is a key parameter to determine the modulating timescale together with the coupling coefficients and linear damping and growth rates. The latter ones cannot be directly obtained from observation while $\delta f$ can be measured if the consecutive subset light curves are long enough. The values of those quantities determine the exact modulation patterns of AMs and FMs. 

We do not provide direct calculations of nonlinear amplitude equations constrained by the observed results since many of the physical quantities are not currently available. The linear growth/damping rates and coupling coefficients need sophisticated seismic models before they can be determined. Once those physical quantities are available, we could constrain the growth/damping rates using the observed modulation periods provided in this analysis. At least we can conclude that  oscillation modes are unstable and this characteristic is a ubiquitous phenomenon in oscillation modes of this pulsating sdB stars. 
 
Other mechanisms can also produce frequency modulations but in a systematic trend. For instance, magnetic cycles generally lead all frequencies to shift with a similar pattern \citep{2015A&A...578A.137S}.  Magnetic field is very rare or completely absent in sdB stars \citep{2012A&A...541A.100L}, except one particular object claimed to be produced through the merger channel \citep{2021A&A...655A..43V}. In addition, sdB stars have very stable radiative envelopes and are not known to show magnetic cycles. Frequency or phase modulations can be induced by orbital companions through periodic variations of light travelling time \citep{2007Natur.449..189S,2018A&A...611A..85S,2015MNRAS.450.4475M}. This kind of FM for all frequencies has to be found with identical orbital period and phase. EPIC~220422705's FMs and AMs cannot be well explained by the above two mechanisms in terms of their modulation patterns. \citet{2020ApJ...890...11M} recently proposed that temporal changes of depth of a surface convective zone can distort the coherent pulsations in hydrogen-atmosphere white dwarfs. That could produce AMs and FMs in sdB stars, but there is no significant convective zone near the surfaces of sdB stars. They also claim a much wider frequency width than what we observed in EPIC~220422705. 

\section{Conclusion}\label{sec:Dis-Conc}

We have analyzed the nearly consecutive {\em K}2 photometry spanning $\sim$76~d on EPIC~220422705, a $g-$mode dominated hybrid pulsating sdB star. A rich frequency spectrum with 66 independent frequencies are detected above the 5.2~$\sigma$ threshold. We attribute 12 frequencies to unresolved rotational multiplets. Rotational periods of $34.04_{-7.12}^{+13.78}$~d, $41.21_{-5.35}^{+7.22}$~d and $28.86_{-7.21}^{+14.34}$~d were derived based on six dipole $g$-modes, three quadruple $g$-modes, and three dipole $p$-modes, respectively. This suggests that EPIC~220422705 has a differential rotation with a slightly  slower core than the envelope. The period spacings within the asymptotic regime are derived with $\sim268.5$\,s and 159.4\,s for dipole  and quadruple modes on average, respectively. We thus identified 9 dipole modes and 13 quadruple modes with eight additional periods that could fit both sequences.  

We characterize 22 significant frequencies with amplitude and frequency modulations.
All those frequencies show clear modulation patterns which are then fitted with simple functions and their uncertainties were tested by \texttt{MCMC} simulations. Most of those AMs and FMs can be fitted with periodic patterns with periods on a timescale of months which is relatively shorter than that found from {\sl Kepler} sdBV stars \citep{2018ApJ...853...98Z}. A notable feature of the modulations we detect is that they exhibit (anti-) correlations between their amplitude and frequency, a similar result to that in {\sl Kepler} sdBV stars. Limited by  the duration of {\sl K}2 photometry, we have not performed any detailed characterization of the relationship between resonant modes of those modulations since the frequency resolution is not precise enough.

To  quantitatively determine whether the discovered modulations are intrinsic or result from two close frequencies, a series of close-frequency simulations were produced and sliding LSPs were compared for each of the 22 frequencies. Only five frequencies are well-represented by two close frequencies. Thus we conclude that 17 frequencies have AMs and FMs which must be intrinsic modulations.

A natural interpretation for such mode variability is the nonlinear mode interactions through resonance \citep[see, e.g.,][]{1995A&A...296..405B}. Depending on the physical quantities in the amplitude equations, resonant modes can have various types of modulation patterns both in amplitude and frequency. We  have expelled other mechanisms  account for our findings although they can also generate AM and FM, for instance, phase variations as depth-of-convective-zone changes \citep{2020ApJ...890...11M}. Finally, our results are the first step to precisely characterize the patterns of mode modulations in sdB stars from {\sl K}2 photometry. Similar to recent results from {\sl Kepler} \citep[e.g.,][]{2021ApJ...921...37Z},  as well as several compact pulsators in the continuous view zones of TESS to be analyzed, these AMs and FMs will open a new avenue to develop nonlinear stellar oscillation theory in the near future.

\section*{Acknowledgements}

We thank an anonymous referee for comments to improve the manuscript and the helpful discussion with Dr. Li Gang, Guo Zhao, Xianfei Zhang and Prof. Wei Xing. We acknowledge the support from the National Natural Science Foundation of China (NSFC) through grants 11833002, 11903005, 12090040 and 12090042. W.Z. is supported by the Fundamental Research Funds for the Central Universities. 
S.C. is supported by the Agence Nationale de la Recherche (ANR, France) under grant ANR-17-CE31-0018, funding the INSIDE project, and financial support from the Centre National d'Études Spatiales (CNES, France). The authors gratefully acknowledge the {\em Kepler} team and all who have contributed to making this mission possible. Funding for the {\em Kepler} mission is provided by NASA's Science Mission Directorate.

\bibliography{reference.bib}{}
\bibliographystyle{aasjournal}

\end{document}

%% file: Fre_n_l.tex
\tablecaption
{All frequencies above the 5.2$\sigma$ threshold detected in EPIC~220422705, by order of increasing frequency.
Column (1) the identification (ID in order of decreasing amplitude), (2) and (3) lists frequencies in $\mu$Hz and  errors, (4) and (5) periods in seconds and errors, (6) and (7) amplitudes in ppm (parts per million) and errors, (8) signal-to-noise ratio (SNR) above the local noise level, (9) - (11) the quantum number identified by the asymptotic regime (see Sec.~2.4), and (12) comments on whether amplitude or frequency modulations were explored or not.}

\tablehead
{\colhead{ID} & \colhead{Frequency} & \colhead{$\sigma$f} & \colhead{Period} & \colhead{$\sigma$P} & \colhead{Amplitude} & \colhead{$\sigma$A}  & \colhead{S/N} & \colhead{$\ell$} & \colhead{$n_{l=1}$} & \colhead{$n_{l=2}$} & \colhead{Comments}  \\
\colhead{} & \colhead{} & \colhead{($\mu$Hz)} & \colhead{($\mu$Hz)} & \colhead{(s)} & \colhead{(s)} 
& \colhead{(ppm)} & \colhead{(ppm)} & \colhead{}& \colhead{} & \colhead{} & \colhead{}} 
\startdata 
$f_{36}$   &    78.513718  &   0.013490  &    12736.627704  &   2.188323  &   48.530  &   7.840  &    6.2  & 1/2  & 41 & 71  & ...  \\ 
$f_{07}$   &    80.593286  &   0.003472  &    12407.981531  &   0.534585  &  190.710  &   7.930  &   24.0  & 2  & -- & 69  &  AFM  \\ 
$f_{47}$   &    81.745528  &   0.015576  &    12233.085076  &   2.330880  &   42.300  &   7.890  &    5.4  & 1/2  & 39 & 68  &  ...  \\ 
$^*f_{18}$   &    89.477815  &   0.007410  &    11175.954576  &   0.925537  &   89.440  &   7.940  &   11.3  & 2  & -- & 61  &  ...  \\ 
$f_{46}$   &    97.679239  &   0.015759  &    10237.590017  &   1.651719  &   42.470  &   8.020  &    5.3  & 2  & -- & 55  &  ...  \\ 
$f_{22}$   &   103.303172  &   0.008272  &     9680.244884  &   0.775105  &   81.070  &   8.030  &   10.1  & 1  & 30 & --  &  ...  \\ 
$f_{44}$   &   105.886042  &   0.015751  &     9444.115392  &   1.404888  &   42.790  &   8.070  &    5.3  & 2  & -- & 50  &  ...  \\ 
$f_{32}$   &   112.125753  &   0.012740  &     8918.557707  &   1.013372  &   52.680  &   8.040  &    6.6  & 1/2  & 27 & 47  &  ...  \\ 
$^*f_{05}$   &   123.087962  &   0.002951  &     8124.271307  &   0.194805  &  225.750  &   7.980  &   28.3  & 1  & 24 & --  &  AFM  \\ 
$f_{26}$   &   123.938687  &   0.010759  &     8068.505692  &   0.700420  &   61.000  &   7.860  &    7.8  & --  & -- & --  &  ...  \\ 
$f_{42}$   &   129.430699  &   0.014990  &     7726.142304  &   0.894816  &   43.060  &   7.730  &    5.6  & --  & -- & --  &  ...  \\ 
$f_{21}$   &   133.290520  &   0.007839  &     7502.409023  &   0.441213  &   81.810  &   7.680  &   10.7  & 1/2  & 22 & 38  &  ...  \\ 
$f_{16}$   &   139.917867  &   0.006231  &     7147.050042  &   0.318290  &  101.460  &   7.570  &   13.4  & 2  & -- & 36  &  AFM  \\ 
$f_{09}$   &   143.418651  &   0.004274  &     6972.593805  &   0.207766  &  146.640  &   7.510  &   19.5  & 1/2  & 20 & 35  &  AFM  \\ 
$f_{28}$   &   153.730694  &   0.010519  &     6504.881834  &   0.445117  &   58.670  &   7.390  &    7.9  & 2  & -- & 32  &  ...  \\ 
$f_{48}$   &   159.361976  &   0.014897  &     6275.022613  &   0.586567  &   40.200  &   7.170  &    5.6  & --  & -- & --  &  ...  \\ 
$f_{12}$   &   166.836670  &   0.005122  &     5993.886121  &   0.184000  &  116.840  &   7.170  &   16.3  & 1/2  & 16 & 29  &  AFM  \\ 
$f_{15}$   &   173.403913  &   0.005551  &     5766.882542  &   0.184615  &  103.960  &   6.910  &   15.0  & --  & -- & --  &  AFM  \\ 
$^*f_{14}$   &   180.391866  &   0.005258  &     5543.487202  &   0.161575  &  106.320  &   6.700  &   15.9  & 1  & 14 &   &  AFM  \\ 
$f_{34}$   &   183.040112  &   0.011008  &     5463.283371  &   0.328564  &   49.530  &   6.530  &    7.6  & --  & $\times$ & --  &  ...  \\ 
$f_{10}$   &   187.474901  &   0.003790  &     5334.047369  &   0.107838  &  141.280  &   6.410  &   22.0  & --  & -- & --  &  AFM  \\ 
$f_{25}$   &   189.829069  &   0.008628  &     5267.897084  &   0.239427  &   62.020  &   6.410  &    9.7  & --  & -- & --  &  ...  \\ 
$f_{40}$   &   191.060597  &   0.012136  &     5233.941562  &   0.332457  &   43.830  &   6.370  &    6.9  & 2  & -- & 24  &  ...  \\ 
$^*f_{41}$   &   198.891984  &   0.012027  &     5027.854728  &   0.304024  &   43.560  &   6.270  &    6.9  & 1  & 12 & --  &  ...  \\ 
$f_{31}$   &   203.922726  &   0.009663  &     4903.818333  &   0.232369  &   53.600  &   6.200  &    8.6  & 2  & -- & 22  &  ...  \\ 
$f_{02}$   &   207.239959  &   0.000919  &     4825.324262  &   0.021389  &  555.800  &   6.120  &   90.9  & --  & -- & --  &  AFM  \\ 
$f_{56}$   &   213.023531  &   0.015460  &     4694.317078  &   0.340689  &   32.410  &   6.000  &    5.4  & --  & -- & --  &  ...  \\ 
$f_{08}$   &   218.362945  &   0.002993  &     4579.531562  &   0.062779  &  166.040  &   5.950  &   27.9  & 2  & -- & 20  &  AFM  \\ 
$f_{57}$   &   237.295070  &   0.015685  &     4214.162557  &   0.278551  &   30.530  &   5.740  &    5.3  & --  & -- & --  &  ...  \\ 
$f_{04}$   &   261.267965  &   0.001401  &     3827.488003  &   0.020527  &  326.870  &   5.490  &   59.6  & 1/2  & 8 & 15  &  AFM  \\ 
$f_{01}$   &   279.767107  &   0.000388  &     3574.401619  &   0.004957  &  1172.450  &   5.450  &  215.2  & --  & $\times$ & --  &  AFM  \\ 
$f_{51}$   &   281.531170  &   0.012426  &     3552.004566  &   0.156770  &   36.490  &   5.430  &    6.7  & --  & -- & --  &  ...  \\ 
$^*f_{23}$   &   288.561685  &   0.005718  &     3465.463543  &   0.068675  &   80.340  &   5.500  &   14.6  & 1  & 7 & --  &  AFM  \\ 
$f_{13}$   &   299.038501  &   0.003969  &     3344.051003  &   0.044386  &  110.410  &   5.250  &   21.0  & 2  & -- & 12  &  AFM  \\ 
$f_{37}$   &   299.821635  &   0.009171  &     3335.316351  &   0.102016  &   47.790  &   5.250  &    9.1  & --  & -- & --  &  ...  \\ 
$f_{06}$   &   306.222893  &   0.002015  &     3265.595173  &   0.021489  &  208.080  &   5.020  &   41.4  & 1  & 6 & --  &  AFM  \\ 
$f_{11}$   &   311.999395  &   0.003205  &     3205.134420  &   0.032923  &  128.280  &   4.920  &   26.1  & --  & -- & --  &  AFM  \\ 
$^*f_{19}$   &   324.165036  &   0.004423  &     3084.848419  &   0.042093  &   89.190  &   4.730  &   18.9  & 1  & 5 & --  &  AFM  \\ 
$f_{03}$   &   328.746189  &   0.000716  &     3041.860354  &   0.006628  &  529.150  &   4.540  &  116.5  & 2  & -- & 10  &  AFM  \\ 
$f_{27}$   &   358.059551  &   0.005609  &     2792.831516  &   0.043749  &   60.530  &   4.070  &   14.9  & --  & -- & --  &  ...  \\ 
$f_{29}$   &   447.460239  &   0.005372  &     2234.835441  &   0.026831  &   56.730  &   3.650  &   15.5  & 1/2  & 2 & 5  &  ...  \\ 
$f_{50}$   &   466.933411  &   0.007938  &     2141.632997  &   0.036406  &   37.600  &   3.570  &   10.5  & --  & -- & --  &  ...  \\ 
$^*f_{30}$   &   501.251468  &   0.005673  &     1995.006626  &   0.022580  &   54.020  &   3.670  &   14.7  & 1  & 1 & --  &  ...  \\ 
$f_{35}$   &   590.545320  &   0.006904  &     1693.350139  &   0.019798  &   48.610  &   4.020  &   12.1  & 1  & 0 & --  &  ...  \\ 
$f_{66}$   &   591.095424  &   0.013287  &     1691.774221  &   0.038028  &   25.300  &   4.030  &    6.3  & --  & -- & --  &  ...  \\ 
$^*f_{54}$   &   622.788529  &   0.010358  &     1605.681470  &   0.026705  &   32.650  &   4.050  &    8.1  & 2  & -- & 1  &  ...  \\ 
$^*f_{24}$   & 697.626978  &   0.004600  &     1433.430803  &   0.009453  &   71.660  &   3.950  &   18.1  & 2  & -- & 0  &  AFM \\ 
$f_{53}$   &   698.271310  &   0.009803  &     1432.108101  &   0.020105  &   33.660  &   3.950  &    8.5  & --  & -- & --  &  ...  \\ 
$f_{55}$   &   698.742459  &   0.010012  &     1431.142457  &   0.020507  &   32.600  &   3.910  &    8.3  & --  & -- & --  &  ...  \\ 
$f_{78}$   &   738.045426  &   0.015618  &     1354.930150  &   0.028672  &   20.240  &   3.790  &    5.3  & --  & -- & --  &  ...  \\ 
$f_{64}$   &   745.027287  &   0.011907  &     1342.232717  &   0.021451  &   26.370  &   3.760  &    7.0  & --  & -- & --  &  ...  \\ 
$f_{61}$   &   835.597738  &   0.012233  &     1196.748093  &   0.017520  &   26.870  &   3.940  &    6.8  & --  & -- & --  &  ...  \\ 
$f_{70}$   &   977.688720  &   0.013094  &     1022.820434  &   0.013699  &   23.660  &   3.710  &    6.4  & --  & -- & --  &  ...  \\ 
$f_{68}$   &  1250.082073  &   0.012890  &      799.947477  &   0.008249  &   24.200  &   3.740  &    6.5  & --  & -- & --  &  ...  \\ 
$f_{59}$   &  1263.268911  &   0.011052  &      791.597095  &   0.006925  &   28.960  &   3.830  &    7.6  & --  & -- & --  &  ...  \\ 
$f_{73}$   &  1280.609422  &   0.014074  &      780.878216  &   0.008582  &   22.930  &   3.860  &    5.9  & --  & -- & --  &  ...  \\ 
$f_{67}$   &  1293.989776  &   0.013128  &      772.803633  &   0.007840  &   24.410  &   3.840  &    6.4  & --  & -- & --  &  ...  \\ 
$f_{63}$   &  1315.516923  &   0.011645  &      760.157458  &   0.006729  &   26.450  &   3.690  &    7.2  & --  & -- & --  &  ...  \\ 
$f_{76}$   &  1339.686713  &   0.014558  &      746.443172  &   0.008111  &   21.080  &   3.680  &    5.7  & --  & -- & --  &  ...  \\ 
$f_{79}$   &  2702.635472  &   0.013530  &      370.009204  &   0.001852  &   20.170  &   3.270  &    6.2  & --  & -- & --  &  ...  \\ 
$f_{17}$   &  2768.529437  &   0.003033  &      361.202589  &   0.000396  &   89.530  &   3.250  &   27.5  & --  & -- & --  &  AFM  \\ 
$f_{20}$   &  2781.657771  &   0.003223  &      359.497854  &   0.000417  &   86.390  &   3.330  &   25.9  & --  & -- & --  &  AFM  \\ 
$f_{69}$   &  2829.666372  &   0.011498  &      353.398553  &   0.001436  &   24.150  &   3.330  &    7.3  & --  & -- & --  &  ...  \\ 
$^*f_{52}$   &  3706.124351  &   0.009301  &      269.823650  &   0.000677  &   34.070  &   3.800  &    9.0  & 1  & -- & --  &  ...  \\ 
$^*f_{39}$   &  3720.773872  &   0.006863  &      268.761294  &   0.000496  &   45.960  &   3.780  &   12.2  & 1  & -- & --  &  AFM  \\ 
$^*f_{62}$   &  3741.962308  &   0.011572  &      267.239464  &   0.000826  &   26.740  &   3.710  &    7.2  & 1  & -- & --  &  ...  \\ 
\hline 
 & & & \multicolumn{6}{c}{Combination Frequencies} & & & \\
\hline
$f_{38}$   &    95.909702  &   0.014504  &    10426.473881  &   1.576715  &   46.290  &   8.040  &    5.8  & --  & -- & -- & $f_{37}-f_{31}$  \\ 
$f_{45}$   &    98.046119  &   0.015753  &    10199.281873  &   1.638675  &   42.570  &   8.030  &    5.3  & --  & -- & -- & $f_{31}-f_{44}$  \\ 
$f_{33}$   &   110.348492  &   0.013488  &     9062.199093  &   1.107641  &   50.150  &   8.100  &    6.2  & --  & -- & -- & $f_3-f_8$  \\ 
$f_{43}$   &   148.317644  &   0.014488  &     6742.286152  &   0.658587  &   42.970  &   7.460  &    5.8  & --  & -- & -- & $f_3-f_{14}$  \\ 
$f_{49}$   &   232.563301  &   0.012359  &     4299.904569  &   0.228515  &   38.320  &   5.670  &    6.8  & --  & -- & -- &  $1/3f_{24}$ \\ 
$f_{59}$   &   326.972087  &   0.014200  &     3058.365043  &   0.132823  &   26.940  &   4.580  &    5.9  & --  & -- & -- & $4f_{47}$  \\ 
$f_{57}$   &   388.249071  &   0.010933  &     2575.666174  &   0.072529  &   29.350  &   3.840  &    7.6  & --  & -- & -- & $3f_{42}$  \\ 
$f_{73}$   &   587.586784  &   0.015314  &     1701.876263  &   0.044356  &   21.950  &   4.030  &    5.5  & --  & -- & -- & $f_{23}+f_{13}$  \\ 
$f_{70}$   &   764.253882  &   0.013222  &     1308.465713  &   0.022638  &   23.550  &   3.730  &    6.3  & --  & -- & -- & $4f_{40}$ \\ 
$f_{76}$   &  1330.731963  &   0.014910  &      751.466131  &   0.008419  &   20.490  &   3.660  &    5.6  & --  & -- & -- & $f_7+f_{68}$  \\ 
$f_{64}$   &  1343.875889  &   0.011808  &      744.116334  &   0.006538  &   25.620  &   3.620  &    7.1  & --  & -- & -- & $f_7+f_{59}$  \\ 
$f_{71}$   &  1374.605015  &   0.012746  &      727.481705  &   0.006746  &   23.050  &   3.520  &    6.6   & --  & -- & --& $f_7+f_{67}$  \\ 
$f_{74}$   &  1396.135693  &   0.013454  &  716.262757  &   0.006902  &   21.530  &   3.470  &    6.2  & --  & -- & -- & $f_7+f_{63}$
\enddata
\tablecomments
{AM/FM/AFM indicates that the frequency has modulation of amplitude~(AM), frequency~(FM) or both~(AFM).
'$\times$' means the mode identified with period spacing but close to the value identified with potential splitting frequencies.
} 

%% file: AFM_Details.tex
\tablecaption{AM/FM detected in EPIC~220422705, sorted by order of increasing frequency.
Column (1) and (2) the ID and frequency taken from Table\,\ref{Tab:freq}, Column (3) the value of correlation between the amplitude and frequency modulations, Column (4) the indication of amplitude modulation (AM) or frequency modulation (FM), Column (5), (6) and (7) the fitting coefficients amplitude (A), period ($T=2\pi/\omega$) and phase ($\phi$) of sinusoidal function if periodic patterns found, (8), (9) and (10) the coefficients of polynomial fitting up to second order, Column (11) the index of fitting function to the modulation pattern and (12) the comments are discussed for the modulation patterns in Section~\ref{close_fre}}

\tablehead{
\colhead {ID} & \colhead {Fre} & \colhead {Corr} & \colhead {AM/FM} & \colhead {A} & \colhead {T=2 $\pi$/$\omega$} & \colhead {$\phi$} & \colhead {a} & \colhead {b} & \colhead {c} & \colhead {Fitting} & Comment \\
\colhead {} & \colhead {($\mu$Hz)} & \colhead {} & \colhead {} & \colhead {(ppm/nHz)} & \colhead {(d)} & \colhead {[0, 2$\pi$)} & \colhead {(10$^{-3}$)} & \colhead {} & \colhead {} & \colhead {} & \colhead {}
}
\startdata
\multirow{2}{*}{$f_{7}$}  & \multirow{2}{*}{$80.5936$}  & \multirow{2}{*}{$-0.35$} & AM  &  $5^{+1}_{-1} $  & $23^{+1}_{-1}$  &  $2.5^{+0.3}_{-0.4}$ &  -- & $ 0.5^{+0.1}_{-0.1}$ & $ 186^{+1}_{-1}$ &  $G_{4}$  &  \multirow{2}{*}{I} \\ 
& & &  FM & $51^{+2}_{-2}$   & $49^{+1}_{-1}$  &  $5.75^{+0.06}_{-0.05}$ & -- &  -- &  -- &  $G_{3}$  &  \\
\multirow{2}{*}{$^*f_{5}$}  & \multirow{2}{*}{123.0877 } &  \multirow{2}{*}{0.86} & AM  &  $276^{+94}_{-82}$  &$101^{+14}_{-14}$  &  $4.5^{+0.1}_{-0.2}$ &  -- &  $7.6^{+3.6}_{-3.5}$ &  $86^{+76}_{-79}$ &  $G_{4}$  &     \multirow{2}{*}{B} \\ 
& & & FM & $40^{+2}_{-2} $  & $40^{+1}_{-1}$  &  $4.28^{+0.1}_{-0.1}$ &  -- &  $-5.1^{+0.1}_{-0.1}$ &  $120^{+22}_{-22}$ &  $G_{4}$  &     \\
\multirow{2}{*}{$f_{16}$} & \multirow{2}{*}{139.9180} & \multirow{2}{*}{0.27} & AM & ${5.1^{+0.7}_{-0.7}}$   & ${31^{+3}_{-3}}$  &  ${1.1^{+0.5}_{-0.5}}$ &  -- &  ${-0.08^{+0.08}_{-0.08}}$ &$ {-112^{+2}_{-2}}$ &  $G_{4}$  &     \multirow{2}{*}{I} \\ 
 &  & & FM &  $11_{-3}^{+2}$   & $16.3_{-0.7}^{+0.7}$  &  $7.6_{-0.5}^{+0.5}$ &  $77_{-14}^{+14}$ &  ${2.9_{-0.7}^{+0.7}}$ &  ${-13_{-10}^{+10}}$ &  $G_{5}$  & \\
\multirow{2}{*}{$f_{9}$} & \multirow{2}{*}{143.4188} & \multirow{2}{*}{$-0.37$} & AM & -- & -- & -- & -- & $-2.0^{+0.03}_{-0.03}$ &  $193^{+1}_{-1}$ & $G_{1}$  &  \multirow{2}{*}{I} \\ 
 & & & FM & $14^{+1}_{-1}$ & $40.2_{-1.4}^{+1.5}$  &  $4.9^{+0.1}_{-0.1}$ &  -- &  -- & $0.002^{+0.895}_{-0.867}$ & $G_{3}$  &   \\ 
\multirow{2}{*}{$f_{12}$} & \multirow{2}{*}{166.8356} & \multirow{2}{*}{0.21} & AM & --   & --  &  -- &  $430^{+2}_{-2}$ &  $1.6^{+0.1}_{-0.1}$ &  $115^{+1}_{-1}$ & $G_{2}$  &  \multirow{2}{*}{B} \\ 
 & & & FM & $14^{+2}_{-2}$   & $29_{-1}^{+1}$  &  $0.3^{+0.2}_{-0.2}$ &  -- &  -- &  -- &  $G_{3}$  &    \\ 
\multirow{2}{*}{$f_{15}$}   & \multirow{2}{*}{173.4053}   & \multirow{2}{*}{0.68} & AM &  $12.2^{+0.6}_{-0.6}$   & $36.1_{-1.4}^{+1.5}$  &  $3.1^{+0.2}_{-0.2}$ &  -- &  $0.8^{+0.1}_{-0.1}$ &  $107^{+1}_{-1}$   &  $G_{4}$  &    \multirow{2}{*}{I} \\ 
 & & & FM & $11^{+2}_{-2}$ & $39.4_{-2.4}^{+2.9}$  &  $5.7^{+0.3}_{-0.3}$ &  -- &  -- &  --  &  $G_{3}$  &   \\ 
\multirow{2}{*}{$^*f_{14}$}   & \multirow{2}{*}{180.3917}   & \multirow{2}{*}{0.73} & AM &  $59_{-31}^{+28}$   & $45.7_{-1.1}^{+1.2}$  &  $5.9_{-2.7}^{+1.75}$ &  -- &  $1.5^{+0.7}_{-0.7}$ &  $97^{+23}_{-20}$  &  $G_{4}$  &    \multirow{2}{*}{I} \\ 
 & & & FM &  $71^{+9}_{-9}$   & $45.7_{-1.1}^{+1.2}$  &  $5.2^{+0.1}_{-0.1}$ &  -- &  -- &  $-53^{+8}_{-7}$  &  $G_{3}$  &   \\ 
\multirow{2}{*}{$f_{10}$}   & \multirow{2}{*}{187.4747}   & \multirow{2}{*}{0.14} & AM &  $4.2^{+0.6}_{-0.6}$   & $30.6_{-2.2}^{+2.6}$  &  $4.9^{+0.4}_{-0.4}$ &  $-20^{+3}_{-3}$ &  $1.5^{+0.1}_{-0.1}$ &  $128^{+2}_{-2}$  &  $G_{5}$  &    \multirow{2}{*}{I} \\ 
 & & & FM & $10^{+2}_{-2}$   & $29.6_{-1.7}^{+2.0}$  &  $2.5^{+0.3}_{-0.3}$ &  -- &  -- &  $1.8^{+1.3}_{-1.3}$  &  $G_{3}$  &   \\ 
\multirow{2}{*}{$f_{2}$}   & \multirow{2}{*}{207.2400}  &  \multirow{2}{*}{$-0.15$} & AM  &  $7^{+1}_{-1}$   & $18.8_{-0.5}^{+0.6}$  &  $0.6^{+0.2}_{-0.3}$ &  -- &  $1.54^{+0.07}_{-0.07}$ &  $524^{+1}_{-1}$  &  $G_{4}$  &    \multirow{2}{*}{B} \\ 
 &  & & FM & $5.9^{+0.4}_{-0.4} $  & $45.6_{-1.4}^{+1.4}$  &  $5.1^{+0.1}_{-0.1}$ &  -- &  -- &  --  &   $G_{3}$  &   \\ 
\multirow{2}{*}{$f_{8}$}   & \multirow{2}{*}{218.3630}   & \multirow{2}{*}{0.6} & AM   &  $10.7^{+0.5}_{-0.5}$   & $33.2_{-0.6}^{+0.7}$  &  $3.6^{+0.1}_{-0.1}$ &  -- &  -- &  $170^{+1}_{-1}$  &  $G_{3}$  &     \multirow{2}{*}{I} \\ 
 & & & FM & $6^{+1}_{-1}$   & $29.4_{-1.3}^{+1.4}$ &  $2.9^{+0.3}_{-0.3}$ & -- &  -- &  --  &  $G_{3}$  &   \\ 
\multirow{2}{*}{$f_{4}$}   & \multirow{2}{*}{261.2680}   &  \multirow{2}{*}{$-0.45$}  & AM & --   & --  &  -- &  -- &  $0.11^{+0.02}_{-0.02}$ &  $322^{+1}_{-1}$   &  $G_{1}$  &    \multirow{2}{*}{I} \\ 
 &  & & FM & $3.3^{+0.4}_{-0.4}$   & $33.9_{-1.7}^{+1.9}$  & $ 3.2^{+0.2}_{-0.2}$ &  -- &  $ -0.12^{+0.04}_{-0.04}$ & $ 3.4^{+0.6}_{-0.6}$   &  $G_{4}$  &  \\ 
\multirow{2}{*}{$f_{1}$}  &  \multirow{2}{*}{279.7670}  & \multirow{2}{*}{$-0.35$} & AM  & $40^{+9}_{-8}$  & $71.4_{-8.5}^{+9.3}$  &  $0.4^{+0.2}_{-0.3}$ &  -- &  $6.2^{+0.5}_{-0.4}$ &  $1027^{+14}_{-15}$  &  $G_{4}$  &    \multirow{2}{*}{I} \\ 
& & & FM &  $0.51_{-0.13}^{0.12}$   & $43.9_{-7.2}^{+11.6}$  &  $1.5^{+0.7}_{-0.7}$ &  -- &  $-0.03^{+0.01}_{-0.01}$ &  $0.6^{+0.3}_{-0.3}$   &   $G_{4}$  \\ 
\multirow{2}{*}{$^*f_{24}$}   & \multirow{2}{*}{288.5611}  & \multirow{2}{*}{$-0.88$} & AM  &  $5.5^{+0.8}_{-0.8}$   & $35.5_{-3.8}^{+4.9}$  &  $4.6^{+0.3}_{-0.3}$ &  -- &  $-0.9^{+0.1}_{-0.1}$ &  $107^{+1}_{-1}$   &  $G_{4}$  &    \multirow{2}{*}{I} \\ 
 & & & FM  &  $12.8^{+2.6}_{-2.6}$   & $29.4_{-3.3}^{+4.4}$  &  $2.2^{+0.5}_{-0.5}$ &  -- &  $4.0^{+0.3}_{-0.3}$ &  $-65^{+5}_{-5}$  &  $G_{4}$  &   \\
\multirow{2}{*}{$f_{13}$}   & \multirow{2}{*}{299.0409}   & \multirow{2}{*}{0.67}& AM   &  --   & -- &  -- &  $-78^{+2}_{-2}$ &  $7.1^{+0.1}_{-0.1}$ &  $17^{+2}_{-2}$   &  $G_{2}$  &    \multirow{2}{*}{B} \\ 
 & & & FM  &  $18^{+1}_{-1}$   & $14.4_{-0.2}^{+0.2}$  &  $5.4^{+0.2}_{-0.2} $&  -- &  $0.9^{+0.1}_{-0.1}$ &  $-24^{+3}_{-3}$ &  $G_{4}$  &    \\ 
\multirow{2}{*}{$f_{6}$}   & \multirow{2}{*}{306.2220}   & \multirow{2}{*}{$-0.82$} & AM  &  --  & -- &  -- &  $-136^{+2}_{-2}$ &  $7.2^{+0.1}_{-0.1}$ &  $166^{+2}_{-2}$  &  $G_{2}$  &    \multirow{2}{*}{I} \\ 
 & & & FM    &  --   & --  & -- &  $43^{+3}_{-3}$ &  $-2.4^{+0.1}_{-0.1}$ &  $25^{+2}_{-2}$  &  $G_{2}$  &   \\ 
\multirow{2}{*}{$f_{11}$}   & \multirow{2}{*}{311.9932}  &  \multirow{2}{*}{0.12} & AM  & --   & --  &  -- &   $-73_{-2}^{+2}$ &  $3.0_{-0.1}^{+0.1}$ &   $131_{-1}^{+1}$  &  $G_{2}$  &  \multirow{2}{*}{I} \\ 
 & & & FM & $30.5^{+1.5}_{-1.5}$ & $44.5_{-0.9}^{+1.0}$ & $3.4^{+0.1}_{-0.1}$ & --   &  --  &  -- &  $G_{3}$  &  \\
\multirow{2}{*}{$^*f_{18}$}   & \multirow{2}{*}{324.1648}   & \multirow{2}{*}{0.74} & AM   &  $24.7^{+0.5}_{-0.5}$   & $69.6_{-2.1}^{+2.3}$  &  $4.5^{+0.1}_{-0.1}$ &  -- &  -- &  $99.4^{+0.6}_{-0.6}$   &  $G_{3}$  &     \multirow{2}{*}{I} \\ 
 & & & FM  & $81^{+2}_{-2}$   & $50.3_{-0.7}^{+0.8}$  &  $4.25^{+0.05}_{-0.05}$ &  -- &  -- &  $3.6^{+1.6}_{-1.5} $  &  $G_{3}$  &  \\ 
\multirow{2}{*}{$f_{3}$}   &  \multirow{2}{*}{328.7461}  &  \multirow{2}{*}{$-0.45$} & AM  &  $8.6^{+0.5}_{-0.5}$   & $14.4_{-0.1}^{+0.1}$  &  $5.2^{+0.1}_{-0.1}$ &  -- &  $-0.91^{+0.05}_{-0.05}$ &  $514^{+1}_{-1}$   &  $G_{4}$  &    \multirow{2}{*}{I} \\
 & & & FM & $6.1^{+0.3}_{-0.3}$   & $20.4_{-0.2}^{+0.2}$  &  $1.4^{+0.1}_{-0.1}$ &  -- &  $-0.08^{+0.02}_{-0.02}$ &  $2.4^{+0.4}_{-0.4}$   & $G_{4}$  &  \\ 
\multirow{2}{*}{$^*f_{24}$}   & \multirow{2}{*}{697.6268}  &  \multirow{2}{*}{0.07} & AM &  --   & --  &  -- &  -- & -- &  -- &        -- & \multirow{2}{*}{I} \\ 
 & & & FM  & --  & --  &  -- &  -- &  -- &  --  &   -- & \\
\multirow{2}{*}{$f_{17}$}   & \multirow{2}{*}{2768.5293}   & \multirow{2}{*}{0.53} &AM   &  --   & -- &  -- &  -- &  $-0.04^{+0.02}_{-0.02}$ &  $90.6^{+0.5}_{-0.5}$  & $G_{1}$  &   \multirow{2}{*}{I} \\ 
 & & & FM  &  $11^{+1}_{-1}$   & $ 40.1_{-2.3}^{+2.6}$  &  $0.6^{+0.2}_{-0.2}$ &  -- &  $-0.4^{+0.1}_{-0.1}$ &  $7.2^{+1.6}_{-1.6}$   &  $G_{4}$  &  \\
\multirow{2}{*}{$f_{20}$}   & \multirow{2}{*}{2781.6577}   &  \multirow{2}{*}{0.19} & AM &  $40.7^{+1.7}_{-1.6}$   & $58.0_{-1.5}^{+1.6}$  &  $4.6^{+0.1}_{-0.1}$ &  -- &  $0.2^{+0.2}_{-0.2}$ &  $84^{+2}_{-2}$  &   $G_{4}$  &    \multirow{2}{*}{B} \\ 
 & & & FM &  $92^{+3}_{-3} $  & $28.0_{-0.3}^{+0.4}$  &  $5.0^{+0.1}_{-0.1}$ &  -- &  $0.32^{+0.01}_{-0.01}$ &  $-17.4^{+0.6}_{-0.6}$   &  $G_{4}$  &  \\ 
\multirow{2}{*}{$^*f_{44}$}   & \multirow{2}{*}{3720.7735}  &  \multirow{2}{*}{0.52}& AM &  $6.1^{+0.5}_{-0.5}$  &$ 16.7_{-0.4}^{+0.4}$  &  $0.10^{+0.16}_{-0.16}$ &  -- &  $-0.7^{+0.1}_{-0.1}$ &  $78^{+1}_{-1}$   &  $G_{4}$  &     \multirow{2}{*}{I} \\ 
 & & & FM &  $26^{+2}_{-3}$   & $26.3_{-1.4}^{+1.6}$  &  $4.6^{+0.2}_{-0.2}$ &  -- & $ -2.0^{+0.3}_{-0.3}$ &  $33^{+4}_{-4}$  & $G_{4}$  & 
\enddata
\tablecomments
{B and I denote the beating effect and intrinsic modulations of amplitude and frequency, respectively.}